\pdfoutput=1
\documentclass[
reprint,
superscriptaddress,
amsmath,
amssymb,
aps,
pra
]{revtex4-2}

\usepackage{graphicx}
\usepackage{dcolumn}
\usepackage{bm}
\usepackage{textcomp} 
\usepackage{siunitx} 

\usepackage[
colorlinks=true,
linkcolor=blue,
citecolor=blue,
urlcolor=blue,
breaklinks=true
]{hyperref} 

\usepackage[all]{hypcap}    
\usepackage{color}
\usepackage{tikz}
\usepackage{ulem} 
\usetikzlibrary{shapes.geometric}

\makeatletter
\renewcommand{\fnum@figure}{\textbf{Fig. \thefigure}}
\makeatother

\newcommand*{\figref}[2][]{
  \hyperref[{fig:#2}]{%
    \ref*{fig:#2}%
    \ifx\\(#1)\\%
    \else
      \,(#1)%
    \fi
  }%
}
\newcommand{\changed}[1]{{#1}}

\newcommand{\bs}[1]{\boldsymbol{#1}}

\newcommand{\mc}[1]{\mathcal{#1}}

\newcommand{\deleted}[1]{}

\newcommand{\graycircle}{\raisebox{0.pt}{\tikz{\node[draw,scale=0.6,circle,white,fill=white!45!black](){};}}}
\newcommand{\blackcircle}{\raisebox{0.2pt}{\tikz{\node[draw,scale=0.6,circle,black,fill=black](){};}}}
\newcommand{\redcircle}{\raisebox{0.2pt}{\tikz{\node[draw,scale=0.6,circle,red,fill=red](){};}}}

\newcommand{\smallredcircle}{\raisebox{0.5pt}{\tikz{\node[draw,scale=0.3,circle,red,fill=red](){};}}}
\newcommand{\greentriangle}{\raisebox{0.8pt}{\tikz{\node[draw=black!60!green,scale=0.2,regular polygon, regular polygon sides=3,fill=black!60!green,rotate=0](){};}}}
\newcommand{\bluetriangle}{\raisebox{0.8pt}{\tikz{\node[draw=blue,scale=0.2,regular polygon, regular polygon sides=3,fill=blue,rotate=180](){};}}}

\newcommand{\opencircle}{\raisebox{0.2pt}{\tikz{\node[draw,scale=0.6,circle,fill=none](){};}}}
\newcommand{\opensquare}{\raisebox{0.2pt}{\tikz{\node[draw,scale=0.8,rectangle, fill=none](){};}}}
\newcommand{\opentriangle}{\raisebox{0.2pt}{\tikz{\node[draw,scale=0.4, regular polygon, regular polygon sides=3, color=blue](){};}}}
\newcommand{\opentriangleud}{\raisebox{0.2pt}{\tikz{\node[draw,scale=0.4, regular polygon, regular polygon sides=3, color=blue,rotate=180](){};}}}

\newcommand{\yellowline}{\raisebox{2pt}{\tikz{\draw[-,black!40!yellow,solid,line width = 1 pt](0,0) -- (3mm,0);}}}
\newcommand{\purpleline}{\raisebox{2pt}{\tikz{\draw[-,black!40!purple,solid,line width = 1 pt](0,0) -- (3mm,0);}}}
\newcommand{\greenline}{\raisebox{2pt}{\tikz{\draw[-,black!40!green,solid,line width = 1 pt](0,0) -- (3mm,0);}}}
\newcommand{\darkgreenline}{\raisebox{2pt}{\tikz{\draw[-,black!40!teal,solid,line width = 1 pt](0,0) -- (3mm,0);}}}
\newcommand{\blueline}{\raisebox{2pt}{\tikz{\draw[-,black!40!blue,solid,line width = 1 pt](0,0) -- (3mm,0);}}}

\def\bx{\bs{x}}

\def\bM{\bs{M}}

\def\bomega{\bs{\omega}}

\def\btau{\bs \tau}

\def\bu{\bs{u}}

\def\bV{\bs{V}}

\def\bF{\bs{F}}
\def\bbf{\bs{f}}

\def\bq{\bs{q}}

\def\bzero{\bs{0}}
\def\btheta{\bs \theta}

\usepackage{times}

\begin{document}

\title{A simple catch: fluctuations enable hydrodynamic trapping of microrollers by obstacles} 

\affiliation{Department of Physics \& Astronomy, Northwestern University, Evanston, IL 60208, USA}
\affiliation{LadHyX, CNRS, Ecole Polytechnique, Institut Polytechnique de Paris, Palaiseau, 91120, France} 
\affiliation{BCAM - Basque Center for Applied Mathematics, Mazarredo 14, E48009 Bilbao, Basque Country - Spain}
\affiliation{Materials Science Division, Argonne National Laboratory, Lemont, IL 60439, USA}
 \affiliation{Current address: Department of Imaging Physics, Delft University of Technology, Lorentzweg 1, 2628 CJ Delft, the Netherlands}
 
\author{Ernest B. van der Wee}
 \email{e.b.vanderwee@tudelft.nl}
 \affiliation{Department of Physics \& Astronomy, Northwestern University, Evanston, IL 60208, USA}
 \affiliation{Current address: Department of Imaging Physics, Delft University of Technology, Lorentzweg 1, 2628 CJ Delft, the Netherlands}

\author{Brendan C. Blackwell}
\affiliation{Department of Physics \& Astronomy, Northwestern University, Evanston, IL 60208, USA}

\author{Florencio Balboa Usabiaga}
\affiliation{BCAM - Basque Center for Applied Mathematics, Mazarredo 14, E48009 Bilbao, Basque Country - Spain}

\author{Andrey Sokolov}
\affiliation{Materials Science Division, Argonne National Laboratory, Lemont, IL 60439, USA}

\author{Isaiah T. Katz}
\affiliation{Department of Physics \& Astronomy, Northwestern University, Evanston, IL 60208, USA}

\author{Blaise Delmotte}
\email{blaise.delmotte@ladhyx.polytechnique.fr}
\affiliation{LadHyX, CNRS, Ecole Polytechnique, Institut Polytechnique de Paris, Palaiseau, 91120, France}

\author{Michelle M. Driscoll}
\email{michelle.driscoll@northwestern.edu}
\affiliation{Department of Physics \& Astronomy, Northwestern University, Evanston, IL 60208, USA}

\date{\today}

\begin{abstract} 

It is known that obstacles can hydrodynamically trap bacteria and synthetic microswimmers in orbits, where the trapping \changed{time} heavily depends on the swimmer flow field and \changed{noise is needed to escape the trap. }
Here, we use experiments and simulations to investigate the trapping of microrollers by obstacles. 
Microrollers \changed{are rotating particles close to a bottom surface which} have a prescribed propulsion direction \changed{imposed by an external rotating magnetic field}. \changed{The} flow field that drives their motion is quite different from previously-studied swimmers. 
\changed{We found that the trapping time} can be controlled by modifying the obstacle \changed{size} or the colloid-obstacle repulsive potential. 
We detail the mechanisms of the trapping and find two remarkable features: the microroller is confined in the wake of the obstacle and, more importantly, it can only enter the trap \textit{with} Brownian motion. 
While noise is usually needed to escape traps in dynamical systems, here we show it is the only means to reach the hydrodynamic attractor.
 
\end{abstract}

\maketitle
\section*{Introduction}

Colloidal-scale swimmers exhibit complex behaviors~\cite{elgeti2015physics,abbott2016active}, such as swarming~\cite{karani2019tuning}, hydro\-dy\-nam\-ically stabilized motile clusters~\cite{driscoll2017unstable}, oscillatory dynamics~\cite{keber2014topology}, and percolating network states~\cite{prymidis2015self}. 
These swimmers can be classified by the flow field they generate, which governs their propulsion as well as their behavior in complex environments, e.g.\ structured landscapes~\cite{bechinger2016active,kos2018elementary,morin2017distortion}.
There is a strong, applications-based interest in microswimmers, as they can be leveraged to advance both microfluidic applications (micromixing, local advective transport, etc.) and drug delivery systems; it is critical to both of these applications to control swimmer transport in a structured environment (e.g., junctions, the blood stream, porous materials)~\cite{tierno2021transport,martinezcalvo2021active}. The motility of these swimmers is coupled to the hydrodynamic flows they generate, and these flows are strongly modified by obstacles, nearby walls, and other structural features.  Thus, in order to learn how to manipulate and guide these microswimmers through more realistic environments, where they will encounter non-trivial geometries, we must develop a framework to understand how these structured environments modify the transport and propulsion of these particles. As a first step to build this understanding, it is important to study a model system: the interaction of a single swimmer with an obstacle~\cite{bechinger2016active,spagnolie2015geometric,takagi2014hydrodynamic,simmchen2016topographical,wykes2017guiding,sipos2015hydrodynamic,das2019colloidal,hoeger2021steric,chaithanya2021wall,Tahaka2022}.

It has been demonstrated that obstacles can be used to guide swimmer trajectories, both deflecting them~\cite{hoeger2021steric}, as well as trapping them in `bound' orbits\changed{~\cite{takagi2014hydrodynamic,simmchen2016topographical,wykes2017guiding,Tahaka2022,ketzetzi2022activity,faundez2022microbial}}.  By manipulating the geometry of these obstacles, one can gain control over both scattering and trapping.  For example, by using pillars of various sizes, approaching bacteria could be scattered at a particular angle~\cite{hoeger2021steric}, or for larger pillars, trapped in an orbit~\cite{sipos2015hydrodynamic,Tahaka2022}.  Similar trapping has been observed in artificial swimmers~\cite{takagi2014hydrodynamic,simmchen2016topographical,ketzetzi2022activity}, and by using more complex geometries, more exotic behaviors, such as directional trapping can be achieved~\cite{wykes2017guiding}. 

The mechanism behind this ensemble of geometry-mediated behaviors is set by the flow field of the microswimmer; this is a hydrodynamic effect.  Many microswimmers have a dipolar flow in the far-field, the direction of the flow classifies them as either `pushers' (\textit{E. Coli}) or `pullers' (alga \textit{Chlamydomonas}).  The scattering and orbital trapping of dipolar swimmers by spherical obstacles was captured in simulations and a semi-analytical far-field hydrodynamics model by Spagnolie et al.~\cite{spagnolie2015geometric}. Their work demonstrated that the trapping \changed{efficiency} of the obstacles was directly set by the swimmer flow field: puller swimmers were trapped by much smaller obstacles than pusher swimmers.  Additionally, they demonstrated that fluctuations by Brownian motion were necessary for a trapped swimmer to exit the bound orbit. 

While swimmers inducing a dipolar flow field are common, there is another class of microswimmers which generate a quite different flow: microrollers, driven by rotation near a boundary~\cite{driscoll2017unstable}. The flow field around a microroller distinctly differs from the dipolar flow fields around more common pusher or puller swimmers~\cite{delmotte2017minimal,bechinger2016active,spagnolie2015geometric,liebchen2021interactions}: there is no fore-aft symmetry and the flow field is not axisymmetric.
Additionally, in the microroller system, the orientation of propulsion does not diffuse, but is prescribed by a rotating field, and can therefore be externally controlled. These rotating particles generate strong flows, which can lead to a tunable and hydrodynamically-mediated attraction between adjacent microrollers~\cite{driscoll2017unstable,martinez2018emergent,delmotte2019hydrodynamically}.  

Dense suspensions of microrollers give rise to interesting collective effects~\cite{driscoll2017unstable,sprinkle2017large,sprinkle2020active,junot2021collective}, such as the formation of hydrodynamically stabilized motile clusters composed of microrollers~\cite{driscoll2017unstable}.  These emergent structures show great promise in the transport of passive species using magnetic fields for microfluidic devices and drug transport, as the magnetic fields used for external control and are non-invasive to the human body~\cite{alapan2020multifunctional}.  

Just as with other kinds of swimmers, to realize the full potential of these systems requires building an understanding of how their transport is modified by a structured environment.  
As the interaction of a single dipolar swimmer with obstacles is very sensitive to (the sign of) its flow field~\cite{spagnolie2015geometric}, we can expect the microroller to similarly exhibit unique interactions with obstacles due to its particular flow field, as well as its prescribed direction. Therefore, studying the interaction of a single microroller with obstacles is needed for our understanding of microroller transport, but will also increase our knowledge of the generalized problem of hydrodynamics-governed interaction of microswimmers with structured environments.

\begin{figure}
 \centering
 \includegraphics[width=0.5\textwidth]{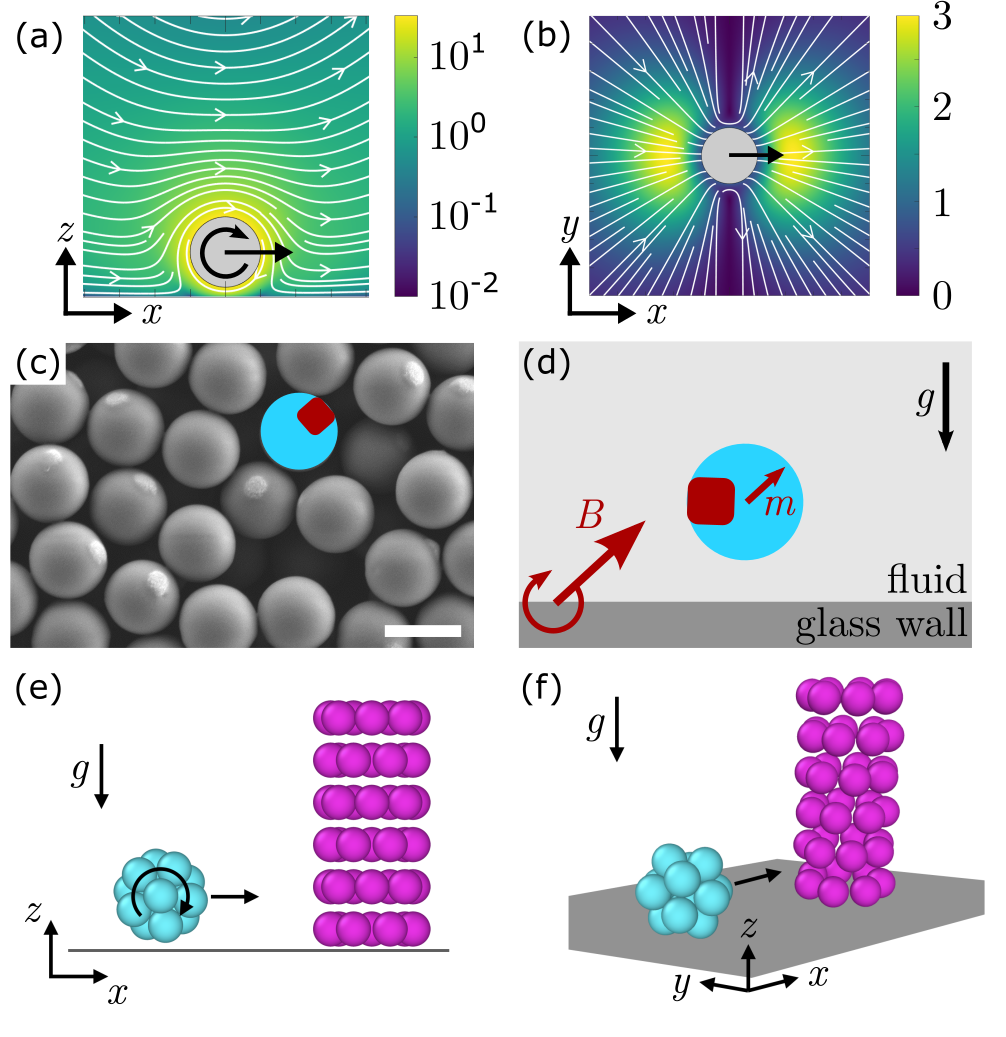}
\caption{
\label{fig:setup}
\textbf{Magnetic field driven microrollers.} 
\changed{The magnitude of velocity (color map) and stream lines (white) of the fluid flow field around a spherical particle rotating perpendicular to a nearby wall in the $x-z$ (a) and $x-y$ plane (b). Velocities are normalized with the bulk velocity of the microroller.}
(c) Scanning electron microscopy image of TPM spheres with an embedded hematite cube and an overlay of a schematic of the particles. The scale bar is 2 \textmu m.
(d) Schematic of a suspended microroller with magnetic moment $m$ confined above a glass wall by gravity $g$ and driven by a magnetic field $B$ rotating perpendicular to the glass wall. 
(e) Side view and (f) perspective view of a microroller (cyan)
with a hydrodynamic radius $r_h$ = 1 \textmu m, constructed as a rigid multiblob, approaching a cylindrical obstacle (magenta) with
a hydrodynamic radius $R_h$ composed of similar sized blobs. The roller is subject to an applied torque in the $x-z$ plane and Brownian motion, whereas the obstacle is frozen into place.}
\end{figure}

Here, we study the interaction of a microroller with a cylindrical obstacle in experiments and through numerical simulations which include Brownian motion and  hydrodynamics~\cite{sprinkle2017large,balboa2017hydrodynamics}.  \changed{The microrollers are rotating colloidal particles confined by a balance between gravity and thermal fluctuations at an average height above a bottom wall~\cite{driscoll2017unstable}.  The (asymmetric) flow field created by the rotation of these microrollers leads to their propulsion (see Figs.~\ref{fig:setup}a-b). 
We note that these particles do not roll on the chamber floor, but are suspended at an average height above it~\cite{driscoll2017unstable}}. This is what allows for such strong hydrodynamic effects in this system: unlike heavier rollers which touch the floor\changed{~\cite{karani2019tuning,chamolly2020irreversible,demirors2021magnetic,bozuyuk2022high,bozuyuk2022reduced}}, the velocity of the fluid at the surface of these  microrollers is orders of magnitude higher than the self-induced velocity of the microrollers themselves (see Figs.~\ref{fig:setup}a-b). In experiments, the microrollers are realized by applying a rotating magnetic field (where the axis of rotation is parallel to the bottom wall) to suspended colloidal particles with a permanent magnetic moment~\cite{driscoll2017unstable,sprinkle2017large,sprinkle2020active,delmotte2017minimal,balboa2017brownian,balboa2017hydrodynamics,junot2021collective}. 

In this system, we observe trapping of the microroller by the obstacle, and demonstrate that this trapping emerges from hydrodynamics alone.  We find that the trapping \changed{time} is sensitive to the relative size of the obstacle, but also depends on the electrostatic repulsion between the obstacle and the microroller; these two control parameters offer unique possibilities for more exotic trapping behaviors.  To understand the mechanism of this trapping, we characterize the \changed{velocity of the roller} around the obstacle and find \changed{saddle points (points of near zero velocity)} up- and downstream of the obstacle, which are connected by a separatrix encircling the obstacle. Near the upstream \changed{saddle point} the roller is repelled from the obstacle, whereas downstream the roller is drawn towards the obstacle, causing it to get trapped \changed{by an attractor (stable node), whose basin of attraction is delimited by the separatrix}. The trapping mechanism we find is quite unique: to enter the basin of attraction of the obstacle, the particle must cross  \changed{the separatrix}.  Thus, in contrast with dipolar swimmers, \changed{noise (e.g., Brownian motion)} is necessary not only to leave the trap, but to enter it as well.  

\section*{Results}
\subsection*{Observations of microroller trapping}
We study the interaction of microrollers with cylindrical obstacles in an experimental system similar to Ref.~\cite{sprinkle2020active}, but with the addition of a 3D-printed cylindrical obstacle. The polymer microrollers with a radius $r = 1.05$ \textmu m contained a hematite cube with a \changed{permanent} magnetic moment (see Fig.~\ref{fig:setup}c). The obstacles were 3D printed on top of a cover glass, from which a sample chamber was built and subsequently filled with a water suspension containing the rollers. We imaged the fluorescently labeled rollers and autofluorescent obstacles using fluorescence microscopy, while applying a rotating magnetic field with the rotation axis parallel to the cover glass (see Fig.~\ref{fig:setup}d).

\begin{figure}
\centering
\includegraphics[width=0.5\textwidth]{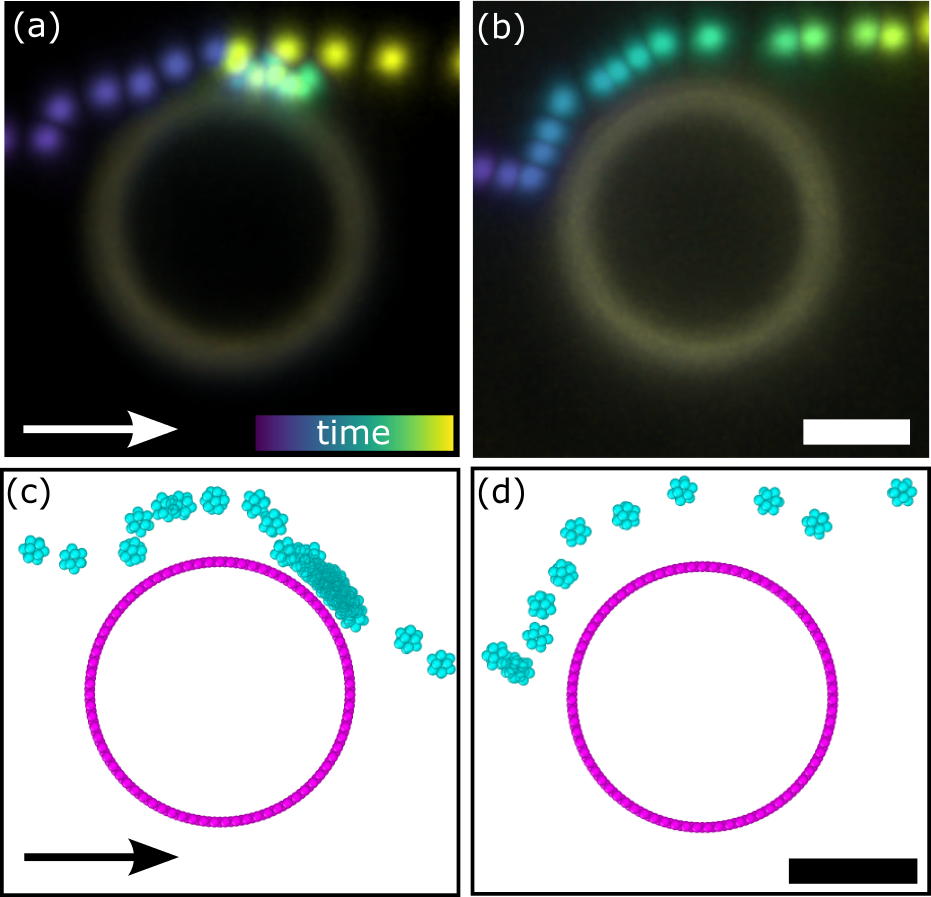}
\caption{
\label{fig:trapping}
\textbf{The interaction of microrollers with cylindrical obstacles in experiments and simulations.}
(a-b) Temporal projections of a fluorescence microscopy image sequence of microrollers interacting with cylindrical obstacles ($H=20$ \textmu m, $R=14.4$ \textmu m), where the microroller is trapped and eventually released (a), or passes the obstacle (b). 
(c-d) Temporal projections of simulations of microrollers approaching cylindrical obstacles ($R_h=10$ \textmu m) where also the microrollers gets trapped and eventually released (c), or passes the obstacle (d).
Videos of the trapping in experiments and simulations are provided in the Supplementary Material~\cite{Note1}.
The arrows denote the direction of propagation of the microrollers.
The scale bars are 10 \textmu m.}
\end{figure}

In Figs.~\ref{fig:trapping}a-b the interaction of a microroller with a printed obstacle with radius $R = 14.4$ \textmu m and height $H = 20$ \textmu m is shown (see also Vids. S1-2 of the Supplementary Material~\footnote{See Supplemental Files at \url{https://doi.org/10.6084/m9.figshare.19772950}}). We observe the trapping and eventual release of the microroller on the side of the roller (see Fig.~\ref{fig:trapping}a, Video S1), but another microroller passes the obstacle without being trapped (see Fig.~\ref{fig:trapping}b, Vid. S2).
 \changed{The} electrostatic interaction between the microrollers and the 3D printed obstacles is purely repulsive as both are negatively charged~\cite{van2017preparation,baker2019shape}, \changed{indicating that} the trapping of the microroller likely originates from hydrodynamics.

We also observed trapping in Brownian dynamics simulations~\cite{sprinkle2017large} of a microroller interacting with an obstacle. The roller and the cylindrical obstacle are modeled as a discrete set of blobs in a coarse-grained model called the rigid multiblob model~\cite{balboa2017hydrodynamics,sprinkle2017large,RigidMultiblobs} (see Figs.~\ref{fig:setup}e-f). A constant torque in the \textit{x-z}-plane is applied to the roller, whereas the obstacle is constrained at a fixed position on the bottom wall.
The microrollers were modeled with a hydrodynamic radius $r_h = $ 1 $\mu$m\cite{geometric} confined by gravity to the no-slip bottom wall. A smaller height of the obstacles ($H = 5.5$ \textmu m)\changed{, with respect to the experiments,} was chosen to reduce the run-time of the simulations and its (minimal) effects on the results will be addressed in the Discussion. The roller, obstacle and wall interacted through a repulsive Yukawa potential. The simulation parameters were chosen similar to the experimental parameters and the work reported in Ref.~\cite{sprinkle2020active}.

We observed hydrodynamic trapping of the roller by the obstacle in our stochastic simulations (see Fig.~\ref{fig:trapping}c and Vid. S3). As in the experiments, the trapping does not always occur, as some rollers pass the obstacles without being trapped (see Fig.~\ref{fig:trapping}d and Vid. S4 of the Supplementary Material~\cite{Note1}).

\subsection*{Microroller interaction with cylindrical obstacles}

\begin{figure*}
\includegraphics[width=\textwidth]{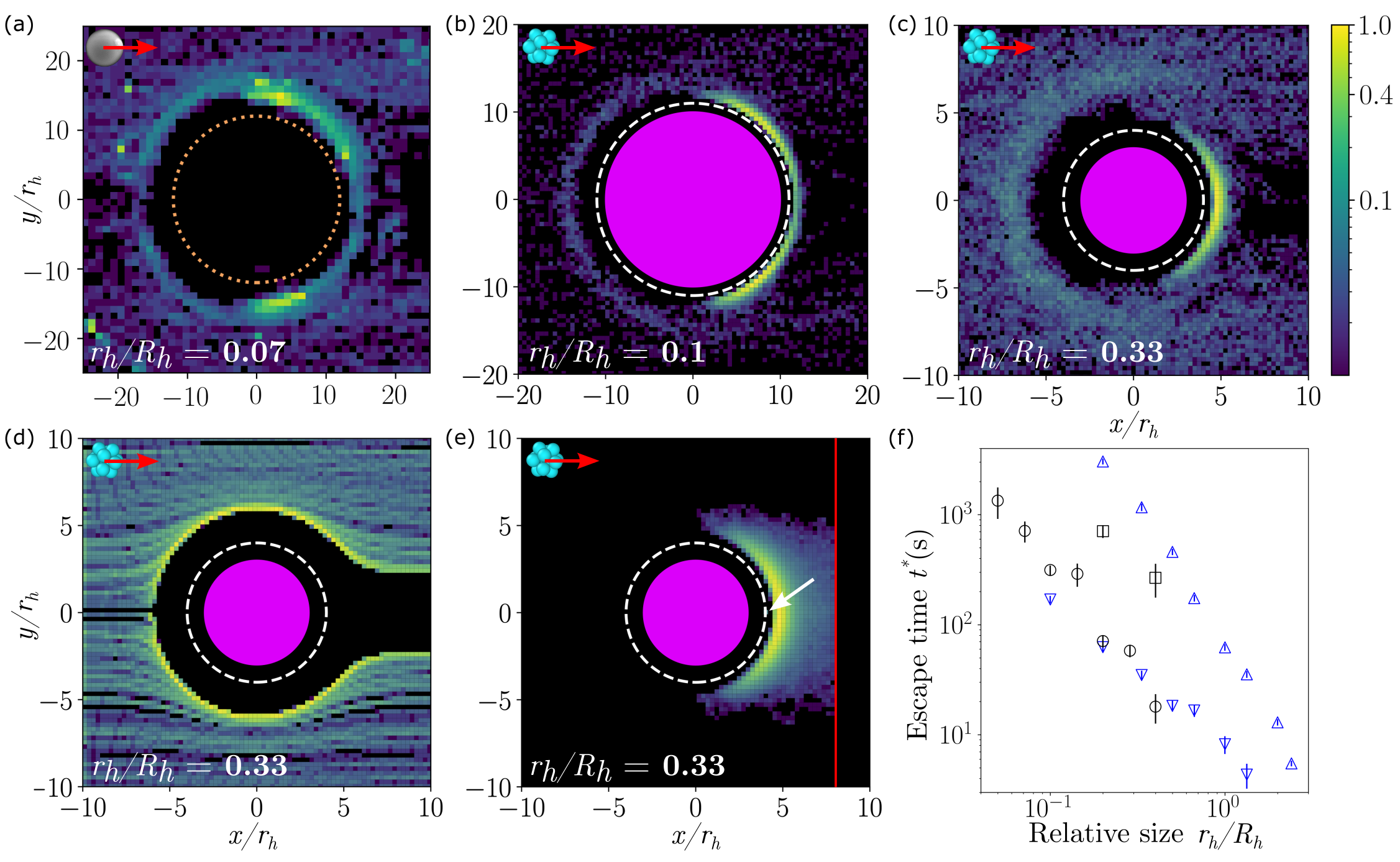}
\caption{
\label{fig:stoch}
\textbf{Microrollers interacting with cylindrical obstacles. }
(a-d) 2D histograms (log-scale) of the [$x$,$y$] coordinates of a microroller interacting with a cylindrical obstacle in experiments (a) and simulations (b-d), for different \changed{relative sizes}: $r_h/R_h = 0.07$ (a), $r_h/R_h = 0.1$ (b) and $r_h/R_h = 0.33$ (c-d). In the simulations, the stochastic (b-c) or deterministic (d) methods were used. The roller is driven in the $x$ direction. In panel (a) the brown dotted circle is drawn as a guide to the eye to clearly show the upstream repulsion and downstream attraction near the obstacle. In panel (b-d), the solid magenta circle denotes the obstacle, the white dashed line the position of the roller at contact with the obstacle. 
(e) 2D histogram (log-scale) of the [$x$,$y$] coordinates of multiple runs where a microroller escapes the hydrodynamic trap of an obstacle ($r_h/R_h = 0.33$) from the starting point at [$x=R_h+r_h$,$y=0$] (white arrow) until the escape when $x>R_h+5r_h$ (red line). Video S5 of the Supplementary Material shows a single escape run~\cite{Note1}.
\changed{The colorbar of the histograms denotes the relative count in log-scale (normalized to the maximum count in the histogram), where zero count values are depicted in black.}
(f) Log-log plot of the mean escape time $t^*$ as a function of \changed{relative size} $r_h/R_h$ in simulations ($b/r_h=0.1$ (\protect\opentriangle) and $b/r_h=0.4$ (\protect\opentriangleud)) and experiments (no salt, $b/r_h \approx 0.3$ (\protect\opencircle) and added salt, $b/r_h = 0.025$ (\protect\opensquare)), where the error bars denote the standard error. \changed{We found exponents of $k = -2.53 \pm 0.08$ (\protect\opentriangle), $-1.33 \pm 0.05$ ( \protect\opentriangleud) and $-1.95 \pm 0.18$ (\protect\opencircle), when fitting the data with $y=ax^k$ (fits not plotted here).}}
\end{figure*}

To study the interaction of microrollers with obstacles in more detail, we measured heat maps (or 2D histograms) of the positions of the microroller around the obstacle, in both experiments (Fig.~\ref{fig:stoch}a) and stochastic simulations (Figs.~\ref{fig:stoch}b-c). In the experiments, we drove microrollers at low area fractions through an array of printed pillars ($r_h/R_h=0.07$) and imaged them by fluorescence microscopy. Using particle tracking~\cite{crocker1996methods,allan2014trackpy}, we assigned the positions of the microrollers to the nearby obstacles and combined this data in a 2D histogram shown in Fig.~\ref{fig:stoch}a. 
Upstream ($x < 0$) a semicircle of low count is observed close to the pillar, indicating a repulsion from the obstacle. Downstream ($x>0$), however, two high count regions are found at about one and five o'clock close to the pillar, indicating an attraction to the obstacle where the roller gets trapped. Furthermore, the hydrodynamic trapping of the particles also results in a low-count zone further downstream of the obstacle. Upstream there is also a lower count around $y=0$, which is caused by the depletion of rollers due the adjacent pillars in the printed array (see Fig. S1 of the Supplementary Material).  

\changed{In the stochastic simulations, 200 runs were performed with the roller at starting positions with $x=-20r_h$ and $y$ ranging from $-10r_h$ to $10r_h$, with steps of $0.1r_h$.} The 2D histograms for $r_h/R_h=0.1$ and $0.33$ are shown in Figs.~\ref{fig:stoch}b-c, respectively. We simulated smaller obstacles than used in the experiments, as the large pillar size used would have led to  long run times due to the number of blobs needed to construct the obstacle in simulations.
Upstream repulsion and downstream attraction are observed, similar to the experiments. 
For $r_h/R_h=0.1$ (see Fig.~\ref{fig:stoch}b), two high count regions are observed, but more downstream than in experiments. For $r_h/R_h=0.33$ (see Fig.~\ref{fig:stoch}c), the two high count regions are merged into a single high count region around $y=0$. 
Similar to the experiments, a depletion zone is found in the wake of the pillar. Furthermore, the width of this depletion zone decreases with increasing \changed{relative size (or relative curvature)} $r_h/R_h$ (see Figs.~\ref{fig:stoch}b-c).  Interestingly, when the simulations are repeated without Brownian motion, using the deterministic Adams-Bashforth method~\cite{balboa2017brownian}, no trapping is observed (see Fig.~\ref{fig:stoch}d).  Instead, the particles are repelled from a low-count zone downstream of the obstacle. This indicates that Brownian motion is needed for the microroller to enter the hydrodynamic trap. 

To investigate the strength of the hydrodynamic trap, we ran stochastic simulations where the particles are placed in the attractive region behind the obstacle at contact [$x=R_h+r_h,y=0$] and the escape time (or first passage time) from the trap is measured (see
Fig.~\ref{fig:stoch}e)\cite{escape}. The escape time is defined as the time it takes the roller to pass $x = R_h + 5r_h$\cite{criterion} (see the red line in
Fig.~\ref{fig:stoch}e). 
The rollers are found to explore the trap by thermal fluctuations and eventually escape (see Vid. S5 in the Supplementary Material~\cite{Note1}).
 The distribution of escape times have a long tail towards longer escape times, as plotted in Fig. S2 of the Supplementary Material. As there is no model describing this process yet, we will resort to using the mean of the distributions to characterize them in further analysis. 

The mean escape time measured in simulations (corrected for the escape time without an obstacle present) $t^* = \langle t_{esc}\rangle - \langle t_{esc}^{no~obstacle} \rangle$\cite{freepassage} as a function of \changed{relative size} $r_h/R_h$ is shown in Fig.~\ref{fig:stoch}f. We ran simulations for two different Debye lengths of the repulsive Yukawa potential \eqref{eq:yukawa}:  $b/r_h=0.1$ (\opentriangle) and $b/r_h=0.4$ (\opentriangleud). We find that the escape time strongly depends on the \changed{relative size} of the obstacle, where \changed{small relative sizes} \changed{lead to long escape times.}
Furthermore, the escape time decreases with an increase in the Debye length and therefore the range of the repulsive electrostatic interaction between the roller and the obstacle. 

To verify these findings, we measured the trapping \changed{time} of cylindrical obstacles in experiments. As we could not place the particles in the wake of the obstacles, we analyzed image sequences of microrollers interacting with cylindrical obstacles and measured the time between a microroller arriving behind the obstacle and it subsequently leaving the trap. Fig.~\ref{fig:stoch}f shows the mean escape time $t^*$ versus \changed{relative size} $r_h/R_h$ data from the experiments for rollers suspended in pure water ($b/r_h \approx 0.3$~\cite{yethiraj2003colloidal}, \opencircle) and in a 0.14 mM LiCl solution ($b/r_h = 0.025$~\cite{sprinkle2020active}, \opensquare).

In the experiments a strong dependence of the mean escape time on the \changed{relative size} is found, similar to the simulations. Moreover, the slopes of the data from the experiments and simulations \changed{are similar}. 
\changed{This is evident from the exponents of $k = -2.53 \pm 0.08$ ($b/r_h=0.1$ (\protect\opentriangle, sim.)), $-1.33 \pm 0.05$ ($b/r_h=0.4$ (\protect\opentriangleud, sim.)), and $-1.95 \pm 0.18$ (pure water, $b/r_h \approx 0.3$ (\protect\opencircle, exp.)), when fitting the data with $y=ax^k$, demonstrating non-linear relations (fits not shown).}
In addition, as in the simulations, an increase in Debye length results in a decrease of the mean escape time. While the escape time versus \changed{relative size} data sets from the simulations and experiments overlap, they do so for different Debye lengths (see Fig.~\ref{fig:stoch}f). Fig.~\ref{fig:stoch}f demonstrates that the trapping \changed{time} can easily be tuned over multiple orders of magnitude in experiments by adjusting both the \changed{relative size} of the obstacle and the Debye length of the microroller suspension. Both of these control parameters are easily accessible experiments by changing the printed obstacle size and/or tuning the salt concentration of the roller suspension.

\subsection*{Basin of attraction }

\begin{figure*}
\includegraphics[width=\textwidth]{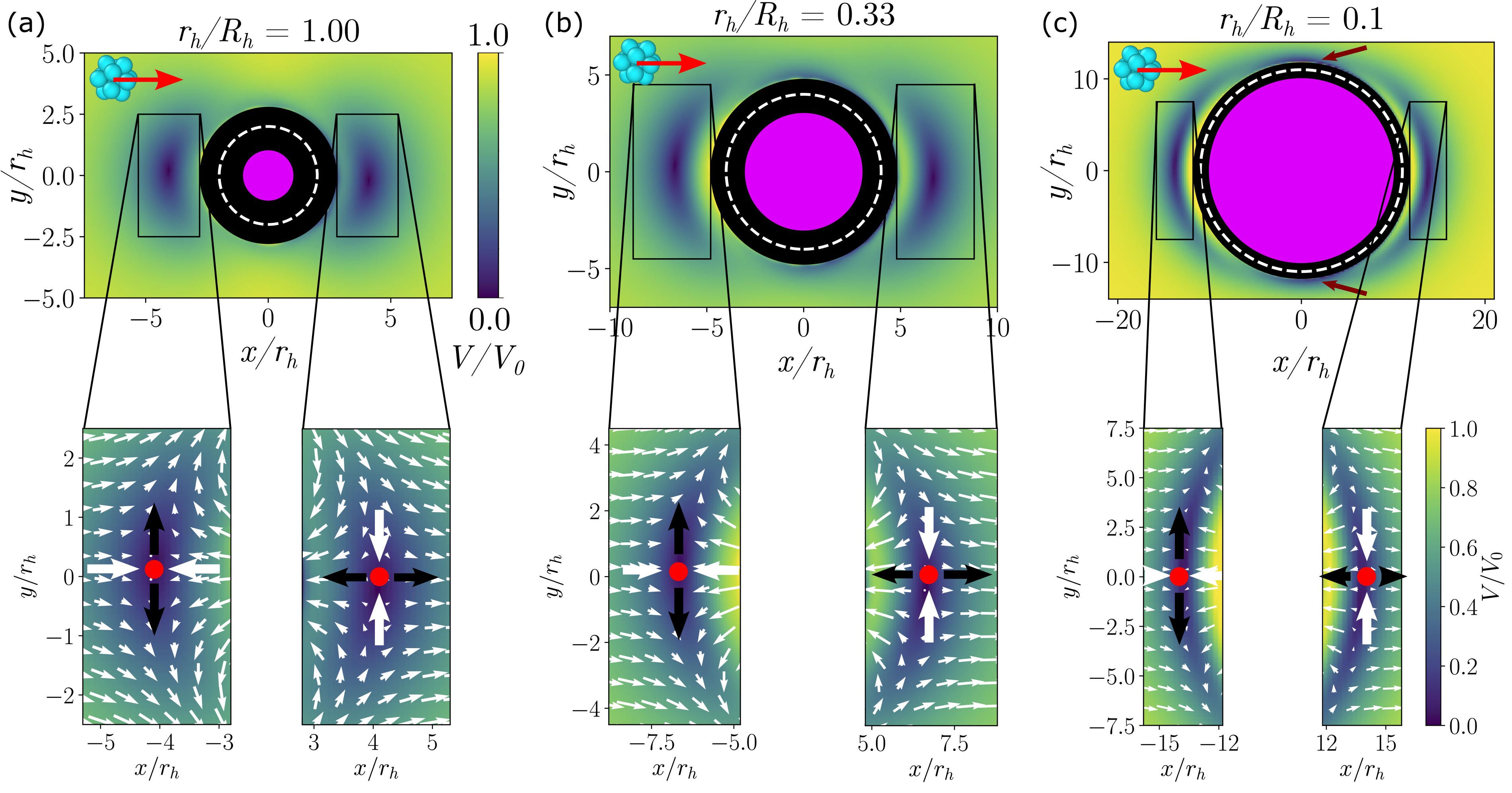}
\caption{
\label{fig:field}
\changed{\textbf{The velocity of a microroller around obstacles of different size.}}
Deterministic microroller velocity fields in the \textit{xy} plane calculated around cylindrical obstacles with \changed{relative sizes of (a) $r_h/R_h = 1.00$, (b) $r_h/R_h = 0.33$ and (c) $r_h/R_h = 0.1$}. \changed{The height of the roller was chosen as the height of the roller in the trap of the obstacle, as determined by simulating a roller placed in the trap using the deterministic method ($z/r_h=$ 1.392, 1.412 and 1.382 for $r_h/R_h =$ 1.0, 0.33 and 0.1, respectively).} The instantaneous microroller velocities in the plots are normalized to the average velocity of the microroller in the absence of obstacles $V_0$. The filled magenta circles denote the obstacle, the white dashed circles the position of the roller at contact with the obstacle. The filled black circle of radius $R_h+r_h+d$, where $d=0.8$, is drawn to block out the region where the electrostatic repulsion dominates the dynamics of the roller. Two \changed{saddle points} ($V/V_0 = 0$, red dots) can be identified upstream and downstream from the obstacle, which differ in their surrounding flow field (white and black arrows): the flow between the point upstream and the obstacle repels the roller from the obstacle, while the point downstream attracts the rollers to the pillar. 
\changed{The brown arrows in (c) indicate the emergence of zero velocity zones for decreasing relative size.}
}
\end{figure*}

To understand the mechanism by which the microrollers are trapped by the obstacles, we calculated the deterministic velocity of the microroller around the obstacles in simulations, allowing us to identify the basin of attraction. This was done by placing the particle on a grid and measuring its instantaneous velocity in the $x-y$ plane at that point.
\changed{The height of the roller was chosen as the height of the roller in the trap of the obstacle, as determined by simulating a roller placed in the trap using the deterministic method.}
The roller velocity fields for \changed{relative size}s $r_h/R_h=1.00$, $r_h/R_h=0.33$ and $r_h/R_h=0.1$ are shown in Fig.~\ref{fig:field}. 
We only plot the roller velocities for $x^2+y^2>(R_h+r_h+d)^2$, where $d=0.8r_h$\cite{electro}, as the microroller velocity too close to the obstacle is dominated by the electrostatic repulsion between the roller and the obstacle.

Two \changed{saddle points (points of near zero velocity) }are identified up- and downstream of the obstacles, as indicated by the red dots in \changed{Figs.~\ref{fig:field}a-c}. Although the two \changed{saddle points} are symmetric with respect to the obstacle, they are tilted slightly with respect to the $x$-axis. This is a non-physical effect induced by the finite resolution of our simulations and the discretized nature of the roller and the obstacle. Although the magnitude of the velocities at the up- and downstream \changed{saddle points} are identical, the directions of the velocities are different (as indicated by the black and white arrows in \changed{Figs.~\ref{fig:field}a-c}): 
while the microroller is pushed from the obstacle between the obstacle and the upstream ($x<0$) \changed{saddle point}, it is pulled into the obstacle downstream ($x>0$).
This explains the regions of low and high count, respectively, up- and downstream of the roller in the 2D histograms in Figs.~\ref{fig:stoch}a-c. 
\changed{
Interestingly, for the roller velocity field of $r_h/R_h = 0.1$, two low velocity regions emerge on the side of the pillar, as indicated by the brown arrows in Fig.~\ref{fig:field}c. These correspond to the two high count regions found in both experiments ($r_h/R_h = 0.07$, Fig.~\ref{fig:stoch}a) and stochastic simulations ($r_h/R_h = 0.1$), Fig.~\ref{fig:stoch}b).}

At a given height, the up- and downstream \changed{saddle points} lie on a \changed{separatrix} forming a circle around the obstacle, where for $x<0$ the flow direction converges at the separatrix, while for $x>0$ the flow diverges (see Fig. S3(b) in the Supplementary Material). This is why in the deterministic simulations of the rollers interacting with the obstacle (see Fig.~\ref{fig:stoch}d) no trapping is observed: as the microroller approaches the \changed{obstacle 
it will never be able to cross the separatrix behind the pillar to reach the 
basin of attraction}. Thus, Brownian motion of the microroller is necessary to cross the separatrix, enter the basin of attraction, and thus be hydrodynamically trapped by the obstacle.

\changed{We find that the saddle point moves non-linearly away from the obstacle surface with decreasing relative size $r_h/R_h$. This effectively increases the size of the basin of attraction, the area where the microroller is attracted to the obstacle (see Figs.~\ref{fig:field}a-c), and results in an increase in trapping time (see Fig.~\ref{fig:stoch}f). As this basin of attraction grows, the fluctuations due to Brownian motion are less likely to kick the roller out of the trap, resulting in longer escape times.
For $r_h/R_h=0$, e.g.\ a wall  ($R_h \to \infty$, 5.5 $\mu$m high, 10 $\mu$m long)  the saddle point becomes a line parallel to the wall 
 (see Fig. S3(c) in the Supplementary Material). 
 } 
 
\begin{figure}
\centering
\includegraphics[width=0.5\textwidth]{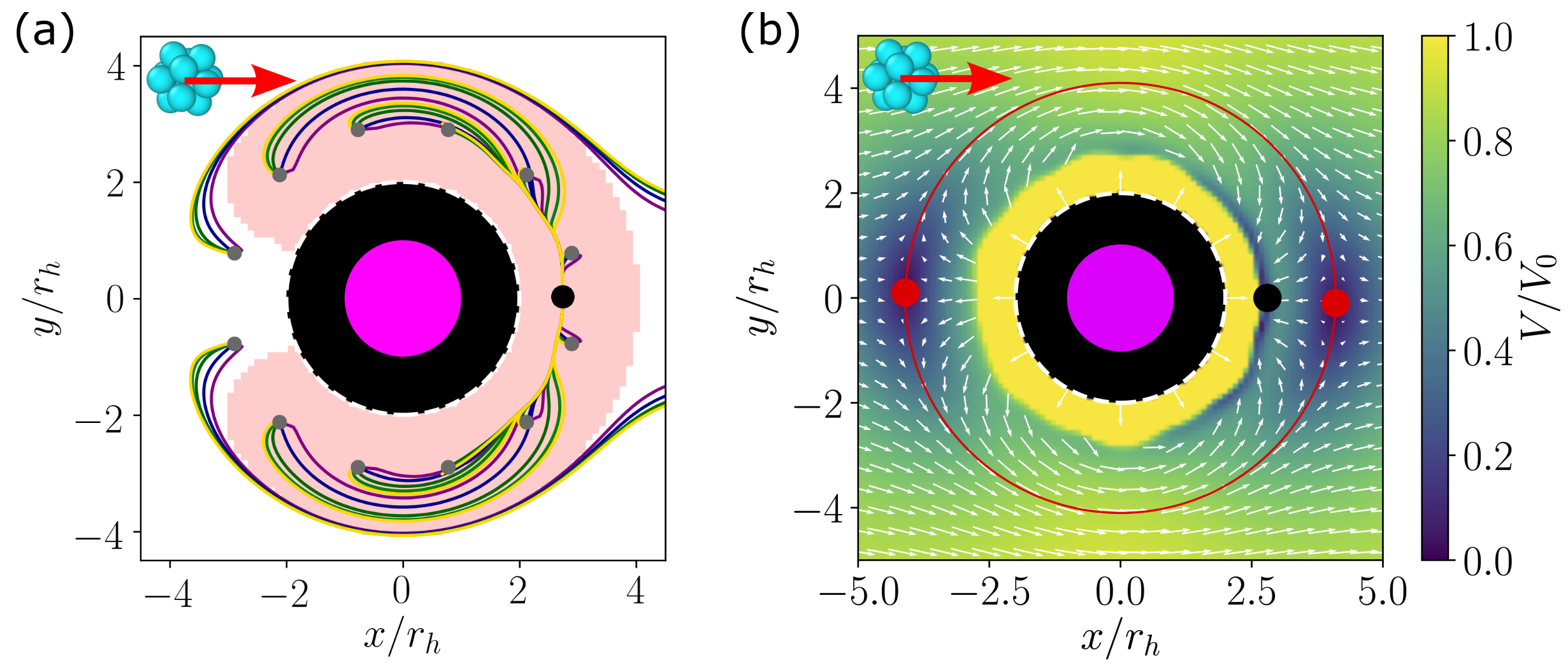}
\caption{
\label{fig:basin}
\changed{\textbf{The basin of attraction and the critical points.}
(a) 2D projection of the trajectories of a roller with different initial positions interacting with a cylindrical obstacle (magenta, $r_h/R_h = 1.00$) calculated using the deterministic Adams-Bashford method. An interactive 3D plot of this panel can be found in File S1 of the Supplementary Material~\cite{Note1}. The initial positions at $x^2+y^2 = 9r_h$ are indicated by \protect\graycircle ~and the trajectories are colored according to the initial roller height $z/r_h$ = 0.9675 (\protect\purpleline), 1.29 (\protect\blueline), 1.935 (\protect\darkgreenline), 2.58 (\protect\greenline) and 3.225 (\protect\yellowline). 
The black and white dashed circle denote the area where the hydrodynamic radius of the roller and obstacle overlap.
The pink area, the basin of attraction, denotes the initial $xy$ positions of rollers with $z=1.392$ that are bound to converge into a single point (\protect\blackcircle) downstream of the obstacle.
This stable node is located on the edge of the zone where the electrostatic repulsion dominates (see black circles drawn in Fig.~\ref{fig:field}a). The fate of the roller (trapping or passing) is independent of the initial height of the roller.
Only the rollers with an initial position outside of the basin of attraction undergo a strong enough hydrodynamic \textit{and} electrostatic repulsion to push the roller around the basin of attraction and past the obstacle. 
(b) The roller velocity field normalized by the bulk velocity $V_0$ as in Fig.~\ref{fig:field}a with the roller velocity directions and critical points annotated. The two saddle points are annotated with \protect\redcircle ~on top of which a circle is plotted indicating the separatrix. The stable node (or attractor) is indicated by \protect\blackcircle. In this plot with also plotted the velocity inside the zone that is dominated by the electrostatic repulsion, where we measured velocities much larger than the bulk velocity $V_0$.
}
}
\end{figure}

In order to calculate the basin of attraction, we ran deterministic simulations for different initial positions around the obstacle and determined whether the microroller got trapped or was able to pass the obstacle. In Fig.~\ref{fig:basin}a the basin of attraction (pink area) around an obstacle ($r_h/R_h=1.00$) is plotted. In addition, we plot the trajectories of rollers with initial positions at $x^2+y^2 = 9r_h$ and different heights, just outside the \changed{area} where the electrostatic repulsion dominates the dynamics of the microroller (see interactive 3D plot in File S1 of the Supplementary Material~\cite{Note1}).
For the majority of initial positions, the roller cannot cross the separatrix and ends up into a \changed{stable node} (as denoted by the black dot in \changed{Fig.~\ref{fig:basin}a}). We find that the roller converges to this point irrespective of its initial height.
Only for initial positions $|y|<0.8, x<0$, which lies outside of the basin of attraction, the hydrodynamic \textit{and} electrostatic repulsion acting on the roller are enough to cross the separatrix and the roller is able to pass the obstacle.

The \changed{stable node} is located on the edge of the area where the roller-pillar interaction is dominated by electrostatic repulsion, as indicated by the black circles in Fig.~\ref{fig:field}\changed{a}. At this point, which is the attractor in this system, the hydrodynamic attraction and electrostatic repulsion acting on the roller are balanced. We can therefore conclude there are three \changed{critical} points in this system \changed{which are summarized in Fig.~\ref{fig:basin}b}: two \changed{saddle} points (up- and downstream) and one attractor or \changed{stable node} (downstream). The two \changed{saddle points} result from the balance between the hydrodynamic interaction (repulsive or attractive) between the roller and the pillar, and the self-induced velocity (or propulsion) of the roller. The \changed{stable node}, however, is a result of the balance between the self-induced velocity of the roller, the hydrodynamic \changed{attraction} \textit{and} the electrostatic repulsion between the roller and the pillar.  \changed{For $r_h/R_h = 1$, the attractor is localized to a single point, and the roller is always trapped immediately behind the obstacle.  For larger obstacles, the situation is more nuanced; regions of near-zero velocity appear not only immediately behind the obstacle, but also in regions along the sides (i.e.\ near 1 o'clock and 5 o'clock., see brown arrows in Fig.~\ref{fig:field}c)}

When the Debye length is increased, the electric repulsion between the roller and the pillar will increase, resulting in a shift of the \changed{stable node} towards the \changed{saddle point}. The reduced distance between the \changed{stable node} and \changed{saddle point}, and therefore the reduced size of the basin of attraction, will increase the probability of the roller leaving the trap due to thermal fluctuations. This agrees with the observed decrease in the measured escape times upon an increase in the Debye length in both experiments and simulations (see Fig.~\ref{fig:stoch}f).

\begin{figure*}
\includegraphics[width=\textwidth]{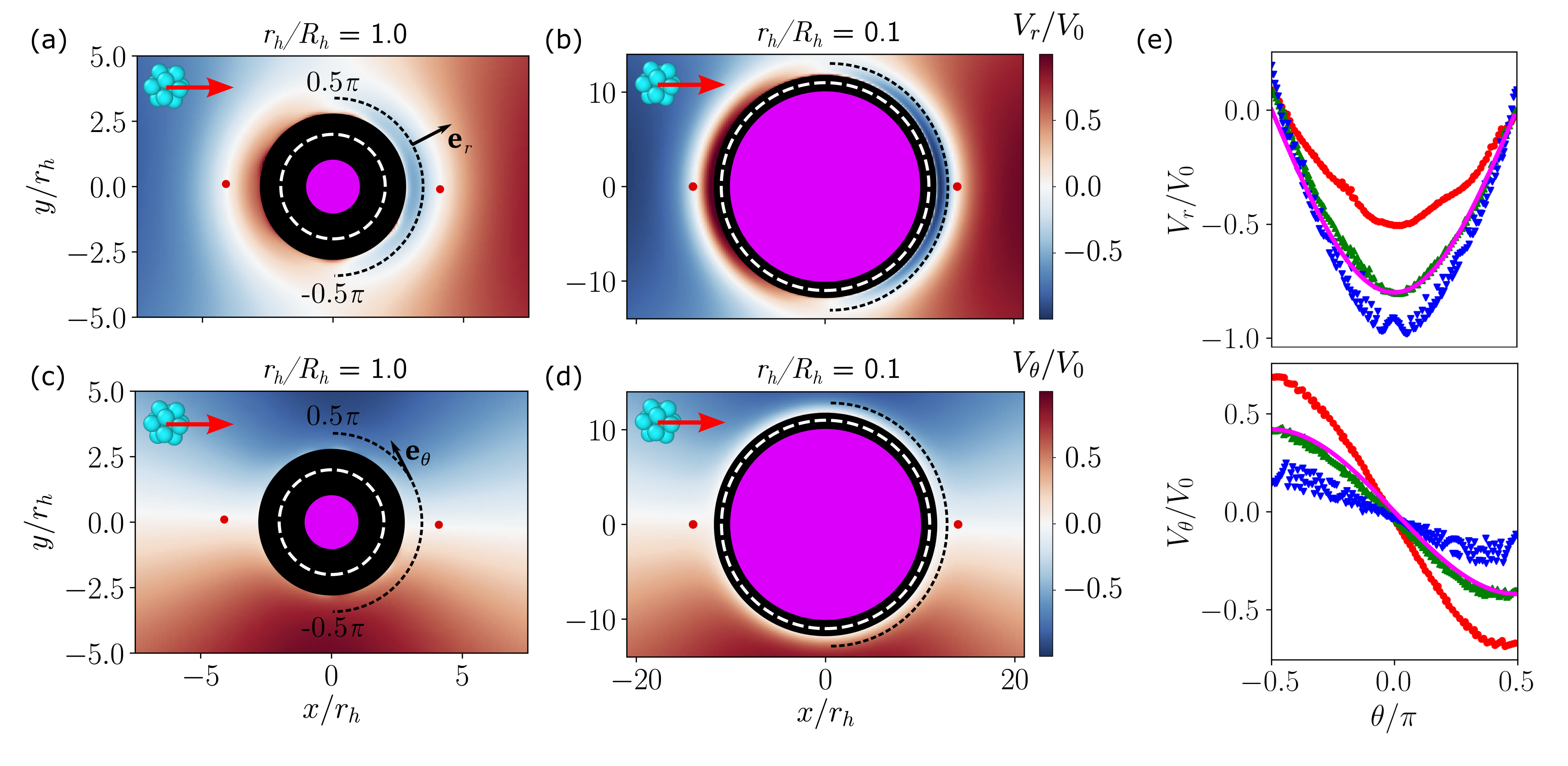}
\caption{
\label{fig:new_figure}
\textbf{Effect of \changed{relative size} on the roller velocity field.}
The radial velocity $V_r$ (a-b) and tangent velocity $V_{\theta}$ (c-d) fields of a roller around obstacles with \changed{relative size} $r_h/R_h = 1.0$ (a,c) and $r_h/R_h=0.1$ (b,d). 
\changed{The velocities were calculated at $z/r_h=$ 1.392 and 1.382 for $r_h/R_h =$ 1.0 and 0.1, respectively.}
The filled magenta circles denote the obstacle, the white dashed circles the position of the roller at contact with the obstacle. The filled black circle of radius $R_h+r_h+d$, where $d=0.8$, is drawn to block out the region where the electrostatic repulsion dominates the dynamics of the roller. The red dots (\protect\smallredcircle) denote the \changed{saddle point} of the roller velocity fields as in Figs.~\ref{fig:field}a,c. 
(e) The radial (top) and tangent (bottom) roller velocities along the black dotted semi-circles in (a-d) as a function of \changed{relative size} $r_h/R_h$ = 1.0 (\protect\smallredcircle), 0.33 (\protect\greentriangle) and 0.1 (\protect\bluetriangle). \changed{The magenta lines correspond to $V_r/V_0 = -0.80 \cos \theta$ (top) and $V_{\theta}/V_0 = -0.42 \sin \theta$ (bottom).}
The fluctuations of the curves for $r_h/R_h =$ 0.1 (\protect\bluetriangle) are due to a relative coarse resolution of the mesh used to calculate the roller velocity fields.
All velocities are normalized with the bulk roller velocity.
}
\end{figure*}

\changed{To further understand how the roller explores the trap geometry,}
we plot the radial and tangent velocity roller velocity fields for \changed{relative sizes} $r_h/R_h = 1.0$ and $0.1$ (see Figs.~\ref{fig:new_figure}a-d). In addition, we plot the radial and tangent velocities along the semi-circles in Figs.~\ref{fig:new_figure}a-d, for \changed{relative sizes} $r_h/R_h=1.0$, $0.33$, and $0.1$ (see Fig.~\ref{fig:new_figure}e). The semi-circles are placed downstream of the pillars, but exactly in between the \changed{saddle point} and \changed{stable node}. \changed{The radial velocity plot (Fig.~\ref{fig:new_figure}e, top) shows that the \changed{relative size} controls the depth of the basin of attraction;  at smaller \changed{relative sizes $r_h/R_h$}, microrollers are more strongly advected to the obstacle, consistent with an increase in escape time in our measurements (Fig.~\ref{fig:stoch}f).} \changed{We also note that the tangent attraction towards the stable node for $r_h/R_h=1.0$ decreases with smaller relative size (see Fig.~\ref{fig:new_figure}e, bottom).} 
Interestingly, the radial and tangent velocities seem to depend on the cosine and sine, respectively, of the angle between the roller-pillar vector and the direction of propulsion (see \changed{magenta} lines in Fig.~\ref{fig:new_figure}e).

\subsection*{Mechanism of microroller trapping}
\begin{figure*}[t]
\includegraphics[width=\textwidth]{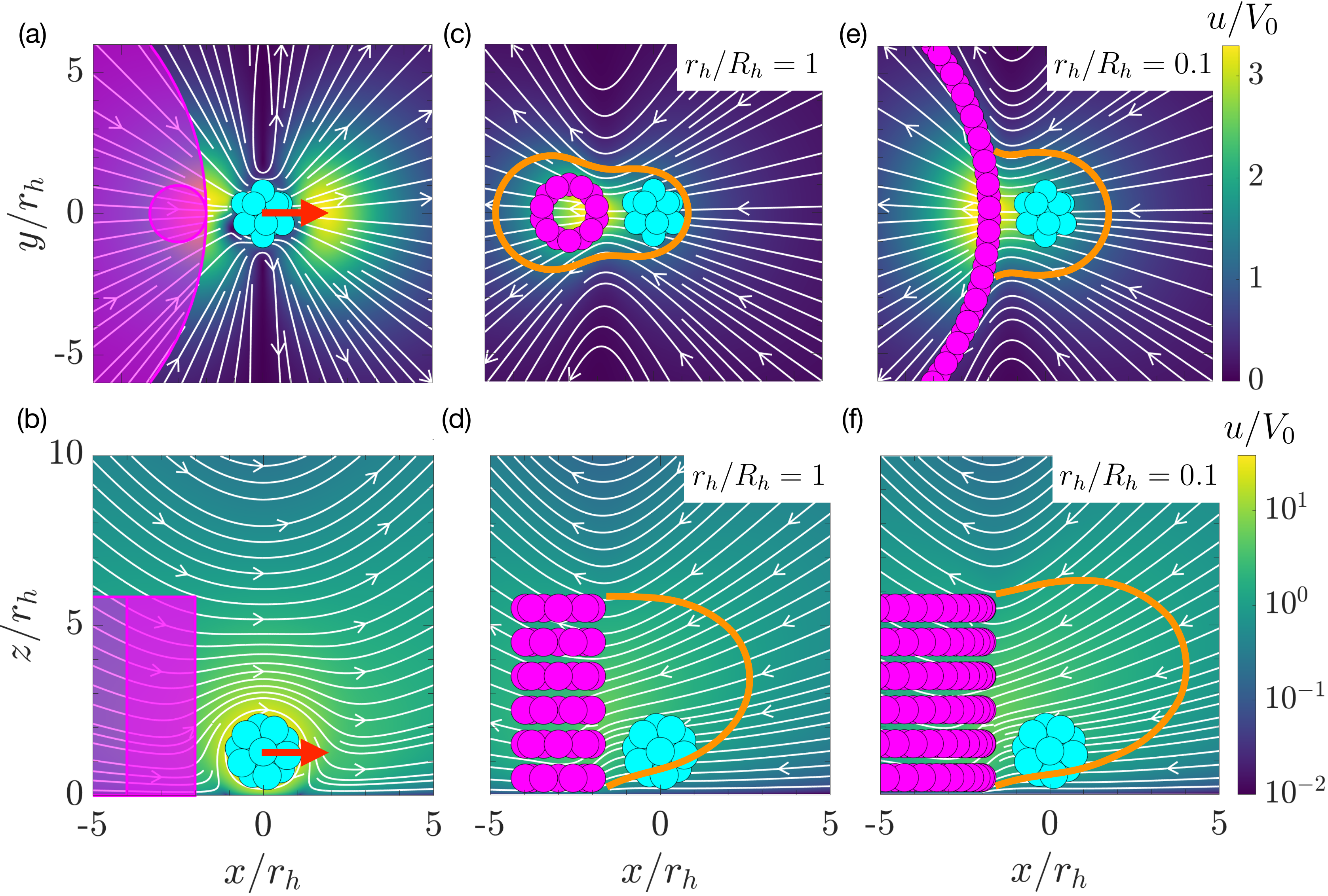}
\caption{\textbf{Trapping mechanism: hydrodynamic response of the obstacle.} Fluid velocity field induced by an isolated roller at a  height $h = 1.29r_h$ in the $x-y$ plane (a) and $x-z$ plane (b). The shaded areas represent the position of an obstacle with \changed{relative size} $r_h/R_h=1$ and $r_h/R_h=0.1$ separated by a horizontal distance $d_x=2r_h$. Streamlines are colored in white and the colorbar represents the magnitude of the flow parallel to the plane and is normalized with the self-induced velocity of a free roller $V_0$. The magnitude of the $x-z$ velocity is shown in log-scale due to the high velocity contrast between the rigid body motion on the roller surface ($u\sim 30V_b$) and the vanishing velocity on the bottom wall ($u=0$). (c)-(f) Fluid velocity field induced by the traction forces on the surface of the cylinder for two \changed{relative sizes} $r_h/R_h=1$ and $r_h/R_h=0.1$ in the $x-y$ plane (c,e) and $x-z$ plane (d,f). Solid orange line: iso-contour $u_x = -V_0$\changed{, for which the fluid velocity induced by the cylinder balances the self-induced velocity of the roller $V_0$. If the microroller lies inside that region, it will be attracted towards the obstacle.  }
\label{fig:flow_obstacle}
}
\end{figure*}

The existence of the \changed{saddle points} has a purely hydrodynamic origin. At the typical height measured in simulations, the flow induced by the microroller is one to two orders of magnitude greater than the self-induced velocity $V_0$: on the roller surface the fluid velocity reaches $u \approx 30V_0$ and $u \approx 5 V_0$ a few radii away along the $x$-axis (see Fig.~\ref{fig:setup}a). As a result, when the obstacle is separated from the microroller at a given distance $d_x$ along the $x$-axis, it needs to cancel strong horizontal (Fig.~\ref{fig:flow_obstacle}a) and vertical (Fig.~\ref{fig:flow_obstacle}b) flows on its surface in order to satisfy the no-slip condition $\bu=0$ for the fluid velocity.  
To do so, it exerts a surface force distribution (called traction forces) that generates a velocity field opposite to the one induced by the microroller (see Figs.~\ref{fig:flow_obstacle}c-f).  Owing to the high magnitude of the surface velocities and to their slow decay at low Reynolds number, the cylinder hydrodynamic response is able to overcome the translation of the microroller at speed $V_0$. This explains why the rollers are attracted to the obstacle at the rear and, by symmetry, repelled at the front. The \changed{saddle points} therefore  correspond to the separation distances at which the cylinder-induced velocity on the microroller exactly balances $V_0$.

Since the area of the cylinder surface exposed to strong flows increases with the cylinder radius $R_h$ (see Figs.~\ref{fig:flow_obstacle}a-b) the reflected flow gets stronger when $r_h/R_h$ decreases and the \changed{saddle points} move away from the cylinder (see Fig.~\ref{fig:field}c).  \changed{The attractive strength of this flow can be measured and }  visualized with the iso-contour  $u_x=-V_0$,  where the horizontal cylinder-induced fluid velocity  balances the microroller velocity, in the $x-y$ and $x-z$ plane (see Figs.~\ref{fig:flow_obstacle}c-f).  \changed{If the microroller lies inside that region, it will be attracted towards the obstacle.} For a fixed horizontal separation distance $d_x=2r_h$, the area of this iso-contour around the microroller increases with $R_h$, leading to an enhanced hydrodynamic attraction. This is further quantified by measuring the volume $\mathcal{V}$ enclosed by the iso-surface $u_x=-V_0$ behind the obstacle: a threefold, \changed{non-linear}, increase of $\mathcal{V}$ is observed between $r_h/R_h=2$ and  $r_h/R_h=0.1$ (see Fig. S4 in the Supplementary Material). 

\begin{figure*}
\includegraphics[width=\textwidth]{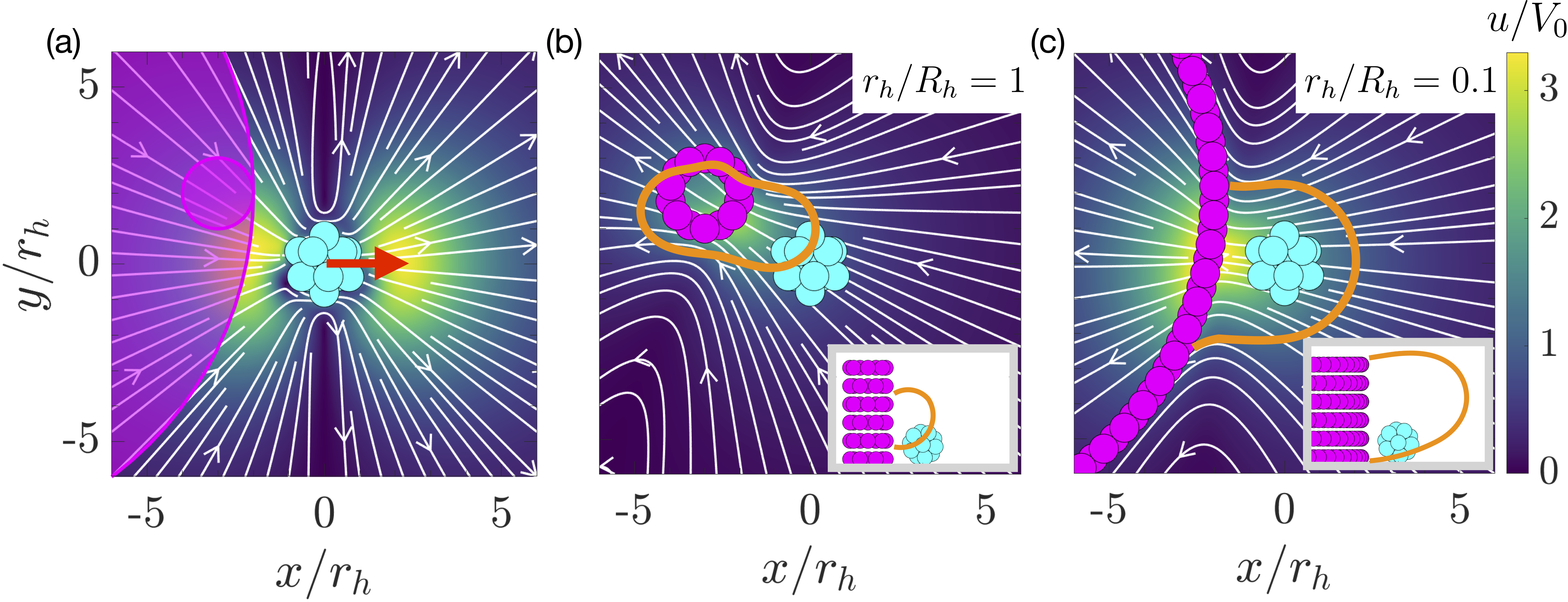}
\caption{ \textbf{Effect of lateral shift on hydrodynamic trapping.} (a) Fluid velocity field induced by an isolated roller at a height $h = 1.29r_h$ in the $xy$-plane. The shaded areas represent the position of an obstacle with \changed{relative size} $r_h/R_h=1$ and $r_h/R_h=0.1$ separated by a horizontal distance $d_x=2r_h$ and lateral distance $d_y = 2r_h$. Streamlines are colored in white and the colorbar represents the magnitude of the flow parallel to the plane and is normalized with the self-induced velocity of a free roller $V_0$. (b)-(c) Fluid velocity field induced by the traction forces on the surface of the cylinder for two \changed{relative sizes} $r_h/R_h=1$ and $r_h/R_h=0.1$ in the $xy$-plane. Solid orange line: iso-contour $u_x = -V_0$. Insets: iso-contour $u_x = -V_0$ in the $xz$-plane going through the center of the microroller at $y=0$.
\label{fig:Flow_obs_dy_2}
}
\end{figure*}

As shown in Figs.~\ref{fig:field}a-b, the cylinder with $r_h/R_h = 0.33$ is able to attract the microroller, i.e.\ induce a negative microroller velocity along the $x$-axis, over a wider range of lateral positions (between $y \approx -2.25r_h$ and $y \approx 2.25r_h$) than for $r_h/R_h = 1$ (between $y \approx -1.25r_h$ and $y \approx 1.25r_h$). Such an increase in the attractive area with $r_h/R_h$ reduces the escape probability from Brownian motion and thus results in longer trapping times.  This increase  can again be explained by looking at the hydrodynamic response of the cylinder surface when it is laterally shifted from the microroller. As shown in Fig.~\ref{fig:Flow_obs_dy_2}a, for a given lateral shift $d_y = d_x = 2r_h$, the magnitude of the flow induced by the microroller at the position of the cylinder surface increases with $R_h$: the larger the cylinder radius, the closer its surface is to the microroller and to the maximal velocity located along the $x$-axis. 
The cylinder response for $d_y = d_x =2r_h$, shown in Figs.~\ref{fig:Flow_obs_dy_2}b-c, is therefore much stronger for $r_h/R_h=0.1$ than for $r_h/R_h=1$: the area of the iso-contour $u_x = -V_0$ does not enclose the microroller anymore for $r_h/R_h=1$, which allows it to escape, while the attractive flow of the largest cylinder is still able to surround  and trap the microroller. 
 In the limit $r_h/R_h\rightarrow 0$, where the cylinder is an infinite wall, the system is translationally invariant along the $y$-axis, the obstacle reflects the microroller flow independently of $d_y$ and the \changed{saddle point} becomes an infinite line (see Fig. S3(c) in the Supplementary Material).

\section*{Discussion}
\label{sec:discussion}
We observed that the escape time versus \changed{relative size} data sets from the simulations and experiments overlap, but for different values of the Debye length (see Fig.~\ref{fig:stoch}f). To match the escape times measured in the experiments, we need to overestimate the Debye length in the simulations. In other words: in simulations a higher trapping \changed{time} is measured than in experiments for the same Debye length.         
We carefully matched the parameters in the simulations, such as the buoyant force, viscosity, and microroller-wall interactions, to the experiments in previous work on dense layers of microrollers~\cite{sprinkle2020active}. As the previous work was in the absence of obstacles,
the simulation parameters concerning the introduced obstacle could very well be the reason of the mismatch. Next, we will discuss the effect of the resolution in our coarse-grained simulations, the height of the pillar and the roller-to-obstacle interaction potential. 

In our simulations the number of blobs per roller of the rigid multiblob model~\cite{balboa2017hydrodynamics} is kept constant at $N=12$. This ensures that the run time of the simulations, where also the obstacles composed of similar sized blobs are present, remains acceptable. It is known, however, that a low resolution in the simulation of a microrollers, leads to an overestimate of its self-induced velocity~\cite{driscoll2017unstable}. Therefore, we also ran simulations with a higher number of blobs per roller ($N=42$) to measure the escape time of the roller for different \changed{relative sizes} $r_h/R_h$ (see Fig. S3a in the Supplementary Material). We find that the escape time of the high resolution roller is increased with respect to the low resolution roller. 
An increase in resolution leads a smaller self-induced velocity while the fluid flow around the roller remains similar, effectively moving the saddle point away from the obstacle, and therefore increasing the escape time~\cite{balboa2017hydrodynamics}.
Instead of narrowing the gap between the measured escape times in the simulations and experiments, this further increases the mismatch.

The height of the pillar $H$ in the experiments was 20 \textmu m, while in the simulations we introduced obstacles with $H=5.5$ \textmu m; this value was chosen to significantly reduce the run-time of the simulations. To study the influence of the pillar height in the simulations, we measured the escape time for different pillar heights with $r_h/R_h=1$ (see Fig. S3(d) in the Supplementary Material). Although the trapping \changed{time} is reduced with a smaller pillar height, at $H=5.5$ \textmu m the escape time is reduced by only $\sim$15$\%$. As the limited pillar heights in the simulations lead to a reduction in the escape times, this cannot explain the larger trapping \changed{time} measured in the simulations.

We have used the same potential at contact $\epsilon$ for the microroller-to-wall (blob-to-wall, $\epsilon_{bw}$) and microroller-to-obstacle (blob-blob, $\epsilon_{bb}$) in the simulations. 
Although we know that the obstacle is negatively charged~\cite{baker2019shape}, just as the glass wall, we do not know the magnitude of the charge and therefore the correct value of $\epsilon_{bb}$. To investigate its influence on the measured escape times, we ran simulations while varying $\epsilon_{bb}$ and keeping $\epsilon_{bw}$ constant. 
In Fig.~S5(a) the mean escape times $t^*$ are plotted for different $\epsilon_{bb}$, simulated with \changed{relative size} $r_h/R_h=1$ and Debye length $(br_h)^{-1}=0.1$. We have also plotted the corresponding interaction potentials in Fig.~S5(b).
The potential at contact $\epsilon_{bb}$ in this work was 0.03, which corresponds to the red point and line in Figs.~S5(a-b).
Clearly, an increase in $\epsilon_{bb}$ leads to a strong reduction of the escape time. 
This reduction can be explained as the screening of the basin of attraction as plotted in Fig.~\ref{fig:field}d, effectively expanding the black circle and making it more probable that the microroller can leave the basin of attraction by Brownian motion.
Therefore, the larger measured escape times in the simulations could very well be explained by an underestimate of the microroller-pillar potential at contact $\epsilon_{bb}$.
One possible way to measure this potential would be by using optical tweezers~\cite{grier1997optical}. Alternatively, the interaction potential can be estimated from zeta potential measurements of colloids fabricated by 3D-printing using the same resin~\cite{baker2019shape,saraswat2020shape,doherty2020catalytically}.

Finally, another possible contribution to the offset in the measured escape times in experiments and simulations could be the different initial conditions in the escape time measurements. Where in the simulations the escape times are measured after placing the roller in the trap in the wake of the obstacle, in the experiments this is not possible and instead the escape times are measured after a roller enters the trap (most often on the side of the obstacle). As pointed out before, placing the roller in the trap in the simulations is done to reduce their run times. Although this might contribute to the offset in the measured escape times, it is unlikely this will change the trend in the escape times as a function of \changed{relative size} as we observe.

\changed{For obstacles that are larger than the roller ($r_h/R_h<1)$, the stable node is no longer a point, but has a more extended geometry consisting of multiple stable nodes.  We believe that this is the origin of the multiple high-count regions in the histograms of experiments and simulations including Brownian motion (Figs.~\ref{fig:stoch}a,b). These trapping points are challenging to resolve precisely in our simulations due to resolution limitations.  The use of high-resolution simulations to study of the precise nature of the changes in the stable node as a function of obstacle geometry would allow a deeper understanding of this dynamical system.  We note that in addition to obstacle geometry, for small relative sizes, the nature of the stable node(s) additionally depends on the rotation frequency (induced flow velocity).  With stronger induced flow, this system becomes fully three dimensional, so that the quasi-2D picture we have used is no longer sufficient to understand the location and geometry of the stable node.  Though it is beyond the scope of this study, understanding this more complex problem is promising avenue for future work.  Exploration along this direction would open up possibilities for finer control of microroller trapping using obstacles with more complex shapes.}

The trapping of active particles has been studied in experiments~\cite{takagi2014hydrodynamic,simmchen2016topographical,wykes2017guiding,sipos2015hydrodynamic,ketzetzi2022activity} and simulations~\cite{spagnolie2015geometric,sharifi2017dynamics}, for bacteria~\cite{sipos2015hydrodynamic,Tahaka2022} and spherical~\cite{takagi2014hydrodynamic,simmchen2016topographical,spagnolie2015geometric,sharifi2017dynamics,ketzetzi2022activity,ketzetzi2022activity} and rod-shaped~\cite{takagi2014hydrodynamic,wykes2017guiding,spagnolie2015geometric} artificial microswimmers. 
The hydrodynamic trapping as reported in these studies is manifested in the orbit of the swimmers around round obstacles and along ridges above a critical \changed{relative size}. 
The escape time of Brownian dipolar swimmers was found to depend on the curvature of the obstacle, as was also put to use in the elegant experiments by Davies-Wykes et al.~\cite{wykes2017guiding}. 
In all of these studies, it was found that Brownian motion only contributes to the \textit{escape} of the swimmers from these orbital traps.
The microrollers studied in our work differ from the swimmers in these studies (\textit{pushers} and \textit{pullers}) by both their flow field~\cite{delmotte2017minimal,kos2018elementary} (see Figs.~\ref{fig:setup}a-b) \textit{and} their restricted orientation as imposed by the plane of rotation of the magnetic field. 
This restriction in the direction of propulsion makes that the trapped microrollers do not orbit the cylindrical obstacle, but rather converge to a single point: the attractor. \changed{This allows for external control of the trapping position by tuning the magnetic field in future applications.}
Moreover, Brownian motion is needed in order for the microrollers to \textit{enter} the basin of attraction, contrary to dipole swimmers, where thermal fluctuations only contribute to the release of the swimmer.

To conclude, we studied the interaction of microrollers with cylindrical obstacles using experiments and simulations including Brownian motion and hydrodynamic interactions. We found hydrodynamic trapping of the rollers downstream to the obstacle, where the trapping \changed{time} increases sharply for \changed{smaller relative size}. The trapping originates from the emergence of a basin of attraction with an attractor behind the obstacle which draws the roller towards the obstacle, which increases with increasing obstacle radius. \changed{At large relative size,} we found three \changed{critical} points of zero roller velocity: two \changed{saddle points} (up- and downstream) and one \changed{stable node} or attractor (downstream). \changed{As relative size is decreased, the stable node transitions from a single point to multiple points, and finally a line in the limit of zero relative size (a wall). } The \changed{saddle points} originate from a balance of the self-induced velocity of the roller and the hydrodynamic interaction with the obstacle, while the attractor adds the electrostatic repulsion between the obstacle and the roller to the balance. Brownian motion plays a double role in the trapping of the microroller: it is needed for the roller to cross a separatrix to enter the trap, but it also kicks the roller out of the trap. This is unlike dipolar microswimmers, such as bacteria (`pushers') and algae (`pullers'), where Brownian motion is only contributing to the escape of the swimmer from the trapped state.
We found an offset in the escape times in simulations and experiments, which we attribute to an underestimate of the obstacle-microroller potential at contact.
Finally, we note that the trapping is easily tunable over orders of magnitude in the laboratory by controlling the curvature of the obstacle and the Debye length of the microroller suspension.

In this work we were limited by the computation time to access higher resolution simulations or larger obstacles ($r_h/R_h \ll 1$), which could be explored in future work, as well as the incorporation of lubrication effects~\cite{sprinkle2020active}. Furthermore, a careful characterization of the microroller-obstacle electrostatic interaction could potentially close the gap in the trapping \changed{time} in the experiments and simulations.

Dense suspensions of microrollers exhibit interesting behavior such as the formation of hydrodynamically stabilized motile clusters~\cite{driscoll2017unstable}. It will be of interest to explore how these motile clusters interact with obstacles, as they are promising for the directed transport of passive cargo~\cite{driscoll2017unstable}. This would be a first step to understand their interaction with a complex environment\changed{, and towards future applications. Preliminary experiments with multiple rollers suggest multiple rollers are less affected by obstacles in their path than single rollers.}

Obstacles with more complex shapes can lead to other interesting hydrodynamic interactions. As shown by Davies Wykes et al.~\cite{wykes2017guiding}, obstacles with a variable curvature (such a teardrops) can lead to controlled release of the swimmer from the obstacle. In the case of the microrollers, this release can then be instigated by a change in the external magnetic field, resulting in switchable and externally controlled trapping. 
Furthermore, as the trapping \changed{time} depends on the \changed{relative size}, it can potentially be used to sort microrollers by their size.
Finally, it is worthwhile to study the interaction of the microrollers with 3D obstacles where the obstacles distort the flow field above the microrollers. It will be interesting to extend this to the interaction of microrollers exploring a 3D environment, for instance porous architectures.

\section*{Materials and Methods}
\subsection*{Experiments}
The experimental system is similar to Ref.~\cite{sprinkle2020active}, but with the addition of a 3D-printed cylindrical obstacle. The microrollers are TPM (3-(trimethoxysilyl)propyl methacrylate) spheres with a diameter of 2.1 \textmu m with an embedded magnetic hematite cube\cite{TPM,sprinkle2020active} (see Figs.~\ref{fig:setup}c-d) suspended in water. To \changed{reduce} the Debye length $b$, lithium chloride (LiCl) was dissolved in the water. 
The obstacles were printed using a photo\-polymer resist (IP-Dip) on microscope cover slips using a Nano\-scribe Professional GT two-photon printer~\cite{nishiguchi2018engineering,reinken2020organizing}. The auto-fluorescent obstacles were printed as open cylinders with height $H= 20$ \textmu m, where the wall thickness was 2.5-2.8 \textmu m, in a periodic array with a square lattice, with a lattice constant of 100 \textmu m.
A sample chamber (${\sim}120$ \textmu m ${\times}~2$ cm~${\times}~2$ cm) was constructed from the cover slip with the printed pillars, two spacers and a microscope slide~\cite{sprinkle2020active}, which was filled with the microroller suspension and sealed using UV glue (Norland Adhesives, no.\ 68). \changed{The glue was cured using UV light, after which the sample was placed on the microscope with the cover slip down. Before imaging the colloids were allowed to sediment toward the cover slip on which the pillars were printed.}

We imaged the microrollers and obstacles using a bright field or fluorescence microscope (see Fig.\ S1 of the Supplementary Material) while applying a rotating magnetic field (40 G, 9 Hz, see Fig.~\ref{fig:setup}d). \changed{The microscope was a Olympus IX83 inverted widefield microscope equipped with a 20$\times$/0.7 NA air objective. 
A 555 nm LED lamp was used for excitation during fluorescence imaging. During the acquisition of long image sequences, the particles were kept in focus using the Olympus IX3-ZDC2 drift compensation module.

The magnetic field was generated using a home-built tri-axial Helmholtz coil set, mounted on top of the microscope stage (see Ref.~\cite{sprinkle2020active} for details and images).
Thorlabs C-mount extension tubes were used to raise the objective close to the sample in the center of the coil set.
To create the rotating magnetic field, two out-of-phase sinusoidal signals were created by a Python script, a data acquisition system (DAQ, Measurement Computing), and two (audio) AC amplifiers (EMB Professional), and fed into the coil set. The phase difference between the two signals was $\pi/2$ and the signals were sent to one coil parallel and one perpendicular to gravity, resulting in a rotating magnetic field with its rotation axis parallel to the bottom wall.

Escape time measurements were done using bright field microscopy and analyzed manually using ImageJ.  Microrollers that remained adjacent to an obstacle after crossing its center coordinate were considered trapped; when these microrollers moved more than 5 particle diameters downstream from the post, they were considered released; the time difference between trapping and release was recorded for multiple interactions (sample size varied between 4-16 escapes per obstacle size).  The times reported represent the mean and standard error of the distribution of trapping times per obstacle size.

2D positional histograms of rollers passing obstacles in experiments were made from two fluorescence microscopy image sequences (sequence length of 1800 frames, captured at 1 frame/second with a 500 ms exposure time) in the following way. 
We tracked the positions of all the rollers using TrackPy~\cite{allan2014trackpy} and filtered out rollers with very short trajectories ($<$$100$ frames), particles that did not move substantially, and clusters of rollers by their intensity. Next, we transformed the particle coordinates so that each particle's position was shifted to the reference frame of the nearest obstacle; obstacle centers were identified using scikit-image's contour finding algorithm \cite{scikit-image}.  To remove instances where two rollers came closer than 15 \textmu m (or 15$r$), we removed these instances from the trajectory as these microrollers likely were hydrodynamically interacting. Finally, all the positions of the remaining rollers around $\sim$30 individual pillars from both image sequences were then combined and plotted in a 2D histogram.
}

\subsection*{Simulations}
To study the trapping of microrollers we performed Brownian dynamics simulations. The dynamics of a microroller satisfy the overdamped Langevin equations \cite{sprinkle2017large}
\begin{equation}
\frac{d\bq}{dt} = \bM \bF + \sqrt{2k_BT}\bM^{1/2}\bs{\mc{W}} + k_BT\partial_{\bq}\cdot\bM
\label{eq:langevin}
\end{equation}
 where $\bq = \{\bx,\btheta\}$ is the vector collecting the roller position $\bx$ and orientation $\btheta$ (here a quaternion). The first term in the RHS of  \eqref{eq:langevin} is the deterministic velocity of the microroller due to the external forces $\bbf$ (here gravity and electrostatic repulsion) and torques $\btau$ (from the rotating magnetic field in experiments)  applied on it, where $\bF=\{\bbf,\btau\}$. The mobility matrix $\bM(\bq(t))$  relates the velocity $\bV$ and rotation rate $\bomega$ to the forces and torques applied on the microroller  through  its hydrodynamic interactions with the wall and the obstacle. The second term is the velocity increment due to  Brownian motion, which involves a vector of independent white noise processes $\bs{\mc{W}}(t)$ and the square root of the mobility matrix $\bM^{1/2}$. The last  term is the stochastic drift  involving the divergence of the mobility matrix with respect to the particle positions and orientations, it  arises when taking the overdamped limit of the Langevin equations \cite{Grassia1996}. More details on the methods used to compute those stochastic terms are found in \cite{sprinkle2017large}.

We computed the mobility matrix $\bM$ by solving the first-kind integral formulation of the Stokes equations with a coarse-grained model called the rigid multiblob model \cite{balboa2017hydrodynamics}, where the continuous single layer potential is replaced by a discrete set of blobs, i.e.\ markers with a finite size, on the surface of the microroller and of the cylinder. These blobs are constrained to satisfy the rigid body motion on the obstacle and microroller surface through a set of Lagrange multipliers.
Hydrodynamic interactions between the blobs are given by a regularization of the Green's function of Stokes equations in the presence of a no-slip wall, called the wall-corrected Rotne-Prager-Yamakawa (RPY) tensor \cite{Swan2007}. The cylinder is constrained at a fixed position on the floor in order to satisfy the no-slip boundary condition for the fluid velocity $\bu=\bzero$ on its surface.

We modeled the microrollers with a hydrodynamic radius $r_h = $ 1 $\mu$m\cite{geometric} confined by gravity to a no-slip bottom wall, while a constant torque is applied in the \textit{x-z}-plane (see Figs.~\ref{fig:setup}e-f)~\cite{balboa2017hydrodynamics,sprinkle2017large,RigidMultiblobs}. 
The microrollers are constructed of 12 blobs with radii $r_b = 0.416~\mu$m, while we vary the hydrodynamics radius $R_h$ of the cylindrical obstacles, which are constructed of blobs with an equal size as the rollers and have a height $H = 5.5$ \textmu m~\footnote{Visualizations throughout this manuscript of the roller and obstacle in simulations were made using OVITO~\cite{ovito}   }. 
A smaller height was chosen in simulations to reduce the run-time of the simulations and its (minimal) effects on the results are addressed in the Discussion. 

The blobs in the roller and pillar interact through a Yukawa potential
\begin{equation}
U(r) =
  \begin{cases}
    \epsilon - \epsilon \frac{r-l}{b}       & \quad \text{if } r < l \\
    \epsilon  e^{-\frac{r-l}{b}}  & \quad \text{if } r \geq l,
  \end{cases}
  \label{eq:yukawa}
\end{equation}
where $\epsilon=0.03$ pN$\mu$m $\approx 7.3$ kT is the repulsion strength at contact, $r$ the center-to-center distance between the blobs, $l$ twice the blob radius $r_b$ and $b/r_h=0.1$ the Debye length. For the interaction between a blob and the bottom wall we use the same potential, but with $l$ equal to the radius of a single blob $r_b$ and $r$ the distance from the wall to the center of the blob~\cite{balboa2017brownian}.
We used the stochastic Trapezoidal Slip method~\cite{sprinkle2017large} to integrate  \eqref{eq:langevin} with a time step $\Delta t / \tau_{self} = 2.25 \times 10^{-3}$, where $\tau_{self}=(6 \pi \eta r_h^3)/k T$, the time the roller takes to diffuse over its own radius in the absence of a driving field~\cite{goodwin2009colloids}. 

The parameters used in the simulations are listed in Tab.~S1 and are chosen similar to the experimental parameters and the work reported in Ref.~\cite{sprinkle2020active}.

\section*{Acknowledgments}
We thank Brennan Sprinkle and Aleksandar Donev for useful discussions, and Mena Youssef and Stefano Sacanna for providing the hematite/TPM particles.
\section*{Funding}
This work was supported by the National Science Foundation under award number CBET-1706562. F.B.U.\ is supported by “la Caixa” Foundation (ID 100010434), fellowship LCF/\-BQ/\-PI20/\-11760014, and from the European Union’s Horizon 2020 research and innovation programme under the Marie Skłodowska-Curie grant agreement No 847648. 
B.D.\ acknowledges support from the French National Research Agency (ANR), under award ANR- 20-CE30-0006. B.D.\ also thanks the NVIDIA Academic Partnership program for providing GPU hardware for performing some of the simulations reported here. The research efforts of A.S. were supported by the U.S. Department of Energy, Office of Science, Basic Energy Sciences, Materials Sciences and Engineering Division.

\section*{Author Contributions}
E.B.W. and M.M.D. conceived and designed research; 
E.B.W., F.B.U. and B.D. performed simulations;
F.B.U. implemented the obstacles into the Rigid Multiblob code;
E.B.W. and I.T.K. built experimental setup;
A.V.S. printed obstacles;
B.C.B., E.B.W. and I.T.K. performed experiments;
E.B.W., B.C.B., B.D. and M.M.D. analyzed data;
E.B.W., B.D. and M.M.D. wrote the manuscript.
   
\section*{Supplementary materials}
\noindent Supporting File Legends\\
Figs. S1 to S5\\
Tab. S1\\
Vids. S1 to S5\\
File S1\\

\bibliography{scifile}

\providecommand{\noopsort}[1]{}\providecommand{\singleletter}[1]{#1}%
\begin{thebibliography}{61}%
\makeatletter
\providecommand \@ifxundefined [1]{%
 \@ifx{#1\undefined}
}%
\providecommand \@ifnum [1]{%
 \ifnum #1\expandafter \@firstoftwo
 \else \expandafter \@secondoftwo
 \fi
}%
\providecommand \@ifx [1]{%
 \ifx #1\expandafter \@firstoftwo
 \else \expandafter \@secondoftwo
 \fi
}%
\providecommand \natexlab [1]{#1}%
\providecommand \enquote  [1]{``#1''}%
\providecommand \bibnamefont  [1]{#1}%
\providecommand \bibfnamefont [1]{#1}%
\providecommand \citenamefont [1]{#1}%
\providecommand \href@noop [0]{\@secondoftwo}%
\providecommand \href [0]{\begingroup \@sanitize@url \@href}%
\providecommand \@href[1]{\@@startlink{#1}\@@href}%
\providecommand \@@href[1]{\endgroup#1\@@endlink}%
\providecommand \@sanitize@url [0]{\catcode `\\12\catcode `\$12\catcode
  `\&12\catcode `\#12\catcode `\^12\catcode `\_12\catcode `\%12\relax}%
\providecommand \@@startlink[1]{}%
\providecommand \@@endlink[0]{}%
\providecommand \url  [0]{\begingroup\@sanitize@url \@url }%
\providecommand \@url [1]{\endgroup\@href {#1}{\urlprefix }}%
\providecommand \urlprefix  [0]{URL }%
\providecommand \Eprint [0]{\href }%
\providecommand \doibase [0]{https://doi.org/}%
\providecommand \selectlanguage [0]{\@gobble}%
\providecommand \bibinfo  [0]{\@secondoftwo}%
\providecommand \bibfield  [0]{\@secondoftwo}%
\providecommand \translation [1]{[#1]}%
\providecommand \BibitemOpen [0]{}%
\providecommand \bibitemStop [0]{}%
\providecommand \bibitemNoStop [0]{.\EOS\space}%
\providecommand \EOS [0]{\spacefactor3000\relax}%
\providecommand \BibitemShut  [1]{\csname bibitem#1\endcsname}%
\let\auto@bib@innerbib\@empty
\bibitem [{\citenamefont {Elgeti}\ \emph {et~al.}(2015)\citenamefont {Elgeti},
  \citenamefont {Winkler},\ and\ \citenamefont {Gompper}}]{elgeti2015physics}%
  \BibitemOpen
  \bibfield  {author} {\bibinfo {author} {\bibfnamefont {J.}~\bibnamefont
  {Elgeti}}, \bibinfo {author} {\bibfnamefont {R.~G.}\ \bibnamefont
  {Winkler}},\ and\ \bibinfo {author} {\bibfnamefont {G.}~\bibnamefont
  {Gompper}},\ }\bibfield  {title} {\bibinfo {title} {Physics of
  microswimmers—single particle motion and collective behavior: a review},\
  }\href@noop {} {\bibfield  {journal} {\bibinfo  {journal} {Reports on
  progress in physics}\ }\textbf {\bibinfo {volume} {78}},\ \bibinfo {pages}
  {056601} (\bibinfo {year} {2015})}\BibitemShut {NoStop}%
\bibitem [{\citenamefont {Abbott}\ and\ \citenamefont
  {Velev}(2016)}]{abbott2016active}%
  \BibitemOpen
  \bibfield  {author} {\bibinfo {author} {\bibfnamefont {N.~L.}\ \bibnamefont
  {Abbott}}\ and\ \bibinfo {author} {\bibfnamefont {O.~D.}\ \bibnamefont
  {Velev}},\ }\bibfield  {title} {\bibinfo {title} {Active particles propelled
  into researchers’ focus},\ }\href@noop {} {\bibfield  {journal} {\bibinfo
  {journal} {Current Opinion in Colloid \& Interface Science}\ }\textbf
  {\bibinfo {volume} {100}},\ \bibinfo {pages} {1} (\bibinfo {year}
  {2016})}\BibitemShut {NoStop}%
\bibitem [{\citenamefont {Karani}\ \emph {et~al.}(2019)\citenamefont {Karani},
  \citenamefont {Pradillo},\ and\ \citenamefont
  {Vlahovska}}]{karani2019tuning}%
  \BibitemOpen
  \bibfield  {author} {\bibinfo {author} {\bibfnamefont {H.}~\bibnamefont
  {Karani}}, \bibinfo {author} {\bibfnamefont {G.~E.}\ \bibnamefont
  {Pradillo}},\ and\ \bibinfo {author} {\bibfnamefont {P.~M.}\ \bibnamefont
  {Vlahovska}},\ }\bibfield  {title} {\bibinfo {title} {Tuning the random walk
  of active colloids: From individual run-and-tumble to dynamic clustering},\
  }\href@noop {} {\bibfield  {journal} {\bibinfo  {journal} {Physical review
  letters}\ }\textbf {\bibinfo {volume} {123}},\ \bibinfo {pages} {208002}
  (\bibinfo {year} {2019})}\BibitemShut {NoStop}%
\bibitem [{\citenamefont {Driscoll}\ \emph {et~al.}(2017)\citenamefont
  {Driscoll}, \citenamefont {Delmotte}, \citenamefont {Youssef}, \citenamefont
  {Sacanna}, \citenamefont {Donev},\ and\ \citenamefont
  {Chaikin}}]{driscoll2017unstable}%
  \BibitemOpen
  \bibfield  {author} {\bibinfo {author} {\bibfnamefont {M.}~\bibnamefont
  {Driscoll}}, \bibinfo {author} {\bibfnamefont {B.}~\bibnamefont {Delmotte}},
  \bibinfo {author} {\bibfnamefont {M.}~\bibnamefont {Youssef}}, \bibinfo
  {author} {\bibfnamefont {S.}~\bibnamefont {Sacanna}}, \bibinfo {author}
  {\bibfnamefont {A.}~\bibnamefont {Donev}},\ and\ \bibinfo {author}
  {\bibfnamefont {P.}~\bibnamefont {Chaikin}},\ }\bibfield  {title} {\bibinfo
  {title} {Unstable fronts and motile structures formed by microrollers},\
  }\href@noop {} {\bibfield  {journal} {\bibinfo  {journal} {Nature Physics}\
  }\textbf {\bibinfo {volume} {13}},\ \bibinfo {pages} {375} (\bibinfo {year}
  {2017})}\BibitemShut {NoStop}%
\bibitem [{\citenamefont {Keber}\ \emph {et~al.}(2014)\citenamefont {Keber},
  \citenamefont {Loiseau}, \citenamefont {Sanchez}, \citenamefont {DeCamp},
  \citenamefont {Giomi}, \citenamefont {Bowick}, \citenamefont {Marchetti},
  \citenamefont {Dogic},\ and\ \citenamefont {Bausch}}]{keber2014topology}%
  \BibitemOpen
  \bibfield  {author} {\bibinfo {author} {\bibfnamefont {F.~C.}\ \bibnamefont
  {Keber}}, \bibinfo {author} {\bibfnamefont {E.}~\bibnamefont {Loiseau}},
  \bibinfo {author} {\bibfnamefont {T.}~\bibnamefont {Sanchez}}, \bibinfo
  {author} {\bibfnamefont {S.~J.}\ \bibnamefont {DeCamp}}, \bibinfo {author}
  {\bibfnamefont {L.}~\bibnamefont {Giomi}}, \bibinfo {author} {\bibfnamefont
  {M.~J.}\ \bibnamefont {Bowick}}, \bibinfo {author} {\bibfnamefont {M.~C.}\
  \bibnamefont {Marchetti}}, \bibinfo {author} {\bibfnamefont {Z.}~\bibnamefont
  {Dogic}},\ and\ \bibinfo {author} {\bibfnamefont {A.~R.}\ \bibnamefont
  {Bausch}},\ }\bibfield  {title} {\bibinfo {title} {Topology and dynamics of
  active nematic vesicles},\ }\href@noop {} {\bibfield  {journal} {\bibinfo
  {journal} {Science}\ }\textbf {\bibinfo {volume} {345}},\ \bibinfo {pages}
  {1135} (\bibinfo {year} {2014})}\BibitemShut {NoStop}%
\bibitem [{\citenamefont {Prymidis}\ \emph {et~al.}(2015)\citenamefont
  {Prymidis}, \citenamefont {Sielcken},\ and\ \citenamefont
  {Filion}}]{prymidis2015self}%
  \BibitemOpen
  \bibfield  {author} {\bibinfo {author} {\bibfnamefont {V.}~\bibnamefont
  {Prymidis}}, \bibinfo {author} {\bibfnamefont {H.}~\bibnamefont {Sielcken}},\
  and\ \bibinfo {author} {\bibfnamefont {L.}~\bibnamefont {Filion}},\
  }\bibfield  {title} {\bibinfo {title} {Self-assembly of active attractive
  spheres},\ }\href@noop {} {\bibfield  {journal} {\bibinfo  {journal} {Soft
  Matter}\ }\textbf {\bibinfo {volume} {11}},\ \bibinfo {pages} {4158}
  (\bibinfo {year} {2015})}\BibitemShut {NoStop}%
\bibitem [{\citenamefont {Bechinger}\ \emph {et~al.}(2016)\citenamefont
  {Bechinger}, \citenamefont {Di~Leonardo}, \citenamefont {L{\"o}wen},
  \citenamefont {Reichhardt}, \citenamefont {Volpe},\ and\ \citenamefont
  {Volpe}}]{bechinger2016active}%
  \BibitemOpen
  \bibfield  {author} {\bibinfo {author} {\bibfnamefont {C.}~\bibnamefont
  {Bechinger}}, \bibinfo {author} {\bibfnamefont {R.}~\bibnamefont
  {Di~Leonardo}}, \bibinfo {author} {\bibfnamefont {H.}~\bibnamefont
  {L{\"o}wen}}, \bibinfo {author} {\bibfnamefont {C.}~\bibnamefont
  {Reichhardt}}, \bibinfo {author} {\bibfnamefont {G.}~\bibnamefont {Volpe}},\
  and\ \bibinfo {author} {\bibfnamefont {G.}~\bibnamefont {Volpe}},\ }\bibfield
   {title} {\bibinfo {title} {Active particles in complex and crowded
  environments},\ }\href@noop {} {\bibfield  {journal} {\bibinfo  {journal}
  {Reviews of Modern Physics}\ }\textbf {\bibinfo {volume} {88}},\ \bibinfo
  {pages} {045006} (\bibinfo {year} {2016})}\BibitemShut {NoStop}%
\bibitem [{\citenamefont {Kos}\ and\ \citenamefont
  {Ravnik}(2018)}]{kos2018elementary}%
  \BibitemOpen
  \bibfield  {author} {\bibinfo {author} {\bibfnamefont {{\v{Z}}.}~\bibnamefont
  {Kos}}\ and\ \bibinfo {author} {\bibfnamefont {M.}~\bibnamefont {Ravnik}},\
  }\bibfield  {title} {\bibinfo {title} {Elementary flow field profiles of
  micro-swimmers in weakly anisotropic nematic fluids: Stokeslet, stresslet,
  rotlet and source flows},\ }\href@noop {} {\bibfield  {journal} {\bibinfo
  {journal} {Fluids}\ }\textbf {\bibinfo {volume} {3}},\ \bibinfo {pages} {15}
  (\bibinfo {year} {2018})}\BibitemShut {NoStop}%
\bibitem [{\citenamefont {Morin}\ \emph {et~al.}(2017)\citenamefont {Morin},
  \citenamefont {Desreumaux}, \citenamefont {Caussin},\ and\ \citenamefont
  {Bartolo}}]{morin2017distortion}%
  \BibitemOpen
  \bibfield  {author} {\bibinfo {author} {\bibfnamefont {A.}~\bibnamefont
  {Morin}}, \bibinfo {author} {\bibfnamefont {N.}~\bibnamefont {Desreumaux}},
  \bibinfo {author} {\bibfnamefont {J.-B.}\ \bibnamefont {Caussin}},\ and\
  \bibinfo {author} {\bibfnamefont {D.}~\bibnamefont {Bartolo}},\ }\bibfield
  {title} {\bibinfo {title} {Distortion and destruction of colloidal flocks in
  disordered environments},\ }\href@noop {} {\bibfield  {journal} {\bibinfo
  {journal} {Nature Physics}\ }\textbf {\bibinfo {volume} {13}},\ \bibinfo
  {pages} {63} (\bibinfo {year} {2017})}\BibitemShut {NoStop}%
\bibitem [{\citenamefont {Tierno}\ and\ \citenamefont
  {Snezhko}(2021)}]{tierno2021transport}%
  \BibitemOpen
  \bibfield  {author} {\bibinfo {author} {\bibfnamefont {P.}~\bibnamefont
  {Tierno}}\ and\ \bibinfo {author} {\bibfnamefont {A.}~\bibnamefont
  {Snezhko}},\ }\bibfield  {title} {\bibinfo {title} {Transport and assembly of
  magnetic surface rotors},\ }\href@noop {} {\bibfield  {journal} {\bibinfo
  {journal} {ChemNanoMat}\ }\textbf {\bibinfo {volume} {7}},\ \bibinfo {pages}
  {881–} (\bibinfo {year} {2021})}\BibitemShut {NoStop}%
\bibitem [{\citenamefont {Martínez-Calvo}\ \emph {et~al.}(2021)\citenamefont
  {Martínez-Calvo}, \citenamefont {Trenado-Yuste},\ and\ \citenamefont
  {Datta}}]{martinezcalvo2021active}%
  \BibitemOpen
  \bibfield  {author} {\bibinfo {author} {\bibfnamefont {A.}~\bibnamefont
  {Martínez-Calvo}}, \bibinfo {author} {\bibfnamefont {C.}~\bibnamefont
  {Trenado-Yuste}},\ and\ \bibinfo {author} {\bibfnamefont {S.~S.}\
  \bibnamefont {Datta}},\ }\bibfield  {title} {\bibinfo {title} {Active
  transport in complex environments},\ }\href@noop {} {\bibfield  {journal}
  {\bibinfo  {journal} {arXiv preprint arXiv:2108.07011}\ } (\bibinfo {year}
  {2021})}\BibitemShut {NoStop}%
\bibitem [{\citenamefont {Spagnolie}\ \emph {et~al.}(2015)\citenamefont
  {Spagnolie}, \citenamefont {Moreno-Flores}, \citenamefont {Bartolo},\ and\
  \citenamefont {Lauga}}]{spagnolie2015geometric}%
  \BibitemOpen
  \bibfield  {author} {\bibinfo {author} {\bibfnamefont {S.~E.}\ \bibnamefont
  {Spagnolie}}, \bibinfo {author} {\bibfnamefont {G.~R.}\ \bibnamefont
  {Moreno-Flores}}, \bibinfo {author} {\bibfnamefont {D.}~\bibnamefont
  {Bartolo}},\ and\ \bibinfo {author} {\bibfnamefont {E.}~\bibnamefont
  {Lauga}},\ }\bibfield  {title} {\bibinfo {title} {Geometric capture and
  escape of a microswimmer colliding with an obstacle},\ }\href@noop {}
  {\bibfield  {journal} {\bibinfo  {journal} {Soft Matter}\ }\textbf {\bibinfo
  {volume} {11}},\ \bibinfo {pages} {3396} (\bibinfo {year}
  {2015})}\BibitemShut {NoStop}%
\bibitem [{\citenamefont {Takagi}\ \emph {et~al.}(2014)\citenamefont {Takagi},
  \citenamefont {Palacci}, \citenamefont {Braunschweig}, \citenamefont
  {Shelley},\ and\ \citenamefont {Zhang}}]{takagi2014hydrodynamic}%
  \BibitemOpen
  \bibfield  {author} {\bibinfo {author} {\bibfnamefont {D.}~\bibnamefont
  {Takagi}}, \bibinfo {author} {\bibfnamefont {J.}~\bibnamefont {Palacci}},
  \bibinfo {author} {\bibfnamefont {A.~B.}\ \bibnamefont {Braunschweig}},
  \bibinfo {author} {\bibfnamefont {M.~J.}\ \bibnamefont {Shelley}},\ and\
  \bibinfo {author} {\bibfnamefont {J.}~\bibnamefont {Zhang}},\ }\bibfield
  {title} {\bibinfo {title} {Hydrodynamic capture of microswimmers into
  sphere-bound orbits},\ }\href@noop {} {\bibfield  {journal} {\bibinfo
  {journal} {Soft Matter}\ }\textbf {\bibinfo {volume} {10}},\ \bibinfo {pages}
  {1784} (\bibinfo {year} {2014})}\BibitemShut {NoStop}%
\bibitem [{\citenamefont {Simmchen}\ \emph {et~al.}(2016)\citenamefont
  {Simmchen}, \citenamefont {Katuri}, \citenamefont {Uspal}, \citenamefont
  {Popescu}, \citenamefont {Tasinkevych},\ and\ \citenamefont
  {S{\'a}nchez}}]{simmchen2016topographical}%
  \BibitemOpen
  \bibfield  {author} {\bibinfo {author} {\bibfnamefont {J.}~\bibnamefont
  {Simmchen}}, \bibinfo {author} {\bibfnamefont {J.}~\bibnamefont {Katuri}},
  \bibinfo {author} {\bibfnamefont {W.~E.}\ \bibnamefont {Uspal}}, \bibinfo
  {author} {\bibfnamefont {M.~N.}\ \bibnamefont {Popescu}}, \bibinfo {author}
  {\bibfnamefont {M.}~\bibnamefont {Tasinkevych}},\ and\ \bibinfo {author}
  {\bibfnamefont {S.}~\bibnamefont {S{\'a}nchez}},\ }\bibfield  {title}
  {\bibinfo {title} {Topographical pathways guide chemical microswimmers},\
  }\href@noop {} {\bibfield  {journal} {\bibinfo  {journal} {Nature
  communications}\ }\textbf {\bibinfo {volume} {7}},\ \bibinfo {pages} {10598}
  (\bibinfo {year} {2016})}\BibitemShut {NoStop}%
\bibitem [{\citenamefont {Wykes}\ \emph {et~al.}(2017)\citenamefont {Wykes},
  \citenamefont {Zhong}, \citenamefont {Tong}, \citenamefont {Adachi},
  \citenamefont {Liu}, \citenamefont {Ristroph}, \citenamefont {Ward},
  \citenamefont {Shelley},\ and\ \citenamefont {Zhang}}]{wykes2017guiding}%
  \BibitemOpen
  \bibfield  {author} {\bibinfo {author} {\bibfnamefont {M.~S.~D.}\
  \bibnamefont {Wykes}}, \bibinfo {author} {\bibfnamefont {X.}~\bibnamefont
  {Zhong}}, \bibinfo {author} {\bibfnamefont {J.}~\bibnamefont {Tong}},
  \bibinfo {author} {\bibfnamefont {T.}~\bibnamefont {Adachi}}, \bibinfo
  {author} {\bibfnamefont {Y.}~\bibnamefont {Liu}}, \bibinfo {author}
  {\bibfnamefont {L.}~\bibnamefont {Ristroph}}, \bibinfo {author}
  {\bibfnamefont {M.~D.}\ \bibnamefont {Ward}}, \bibinfo {author}
  {\bibfnamefont {M.~J.}\ \bibnamefont {Shelley}},\ and\ \bibinfo {author}
  {\bibfnamefont {J.}~\bibnamefont {Zhang}},\ }\bibfield  {title} {\bibinfo
  {title} {Guiding microscale swimmers using teardrop-shaped posts},\
  }\href@noop {} {\bibfield  {journal} {\bibinfo  {journal} {Soft Matter}\
  }\textbf {\bibinfo {volume} {13}},\ \bibinfo {pages} {4681} (\bibinfo {year}
  {2017})}\BibitemShut {NoStop}%
\bibitem [{\citenamefont {Sipos}\ \emph {et~al.}(2015)\citenamefont {Sipos},
  \citenamefont {Nagy}, \citenamefont {Di~Leonardo},\ and\ \citenamefont
  {Galajda}}]{sipos2015hydrodynamic}%
  \BibitemOpen
  \bibfield  {author} {\bibinfo {author} {\bibfnamefont {O.}~\bibnamefont
  {Sipos}}, \bibinfo {author} {\bibfnamefont {K.}~\bibnamefont {Nagy}},
  \bibinfo {author} {\bibfnamefont {R.}~\bibnamefont {Di~Leonardo}},\ and\
  \bibinfo {author} {\bibfnamefont {P.}~\bibnamefont {Galajda}},\ }\bibfield
  {title} {\bibinfo {title} {Hydrodynamic trapping of swimming bacteria by
  convex walls},\ }\href@noop {} {\bibfield  {journal} {\bibinfo  {journal}
  {Physical review letters}\ }\textbf {\bibinfo {volume} {114}},\ \bibinfo
  {pages} {258104} (\bibinfo {year} {2015})}\BibitemShut {NoStop}%
\bibitem [{\citenamefont {Das}\ and\ \citenamefont
  {Cacciuto}(2019)}]{das2019colloidal}%
  \BibitemOpen
  \bibfield  {author} {\bibinfo {author} {\bibfnamefont {S.}~\bibnamefont
  {Das}}\ and\ \bibinfo {author} {\bibfnamefont {A.}~\bibnamefont {Cacciuto}},\
  }\bibfield  {title} {\bibinfo {title} {Colloidal swimmers near curved and
  structured walls},\ }\href@noop {} {\bibfield  {journal} {\bibinfo  {journal}
  {Soft matter}\ }\textbf {\bibinfo {volume} {15}},\ \bibinfo {pages} {8290}
  (\bibinfo {year} {2019})}\BibitemShut {NoStop}%
\bibitem [{\citenamefont {Hoeger}\ and\ \citenamefont
  {Ursell}(2021)}]{hoeger2021steric}%
  \BibitemOpen
  \bibfield  {author} {\bibinfo {author} {\bibfnamefont {K.}~\bibnamefont
  {Hoeger}}\ and\ \bibinfo {author} {\bibfnamefont {T.}~\bibnamefont
  {Ursell}},\ }\bibfield  {title} {\bibinfo {title} {Steric scattering of
  rod-like swimmers in low reynolds number environments},\ }\href@noop {}
  {\bibfield  {journal} {\bibinfo  {journal} {Soft Matter}\ }\textbf {\bibinfo
  {volume} {17}},\ \bibinfo {pages} {2479} (\bibinfo {year}
  {2021})}\BibitemShut {NoStop}%
\bibitem [{\citenamefont {Chaithanya}\ and\ \citenamefont
  {Thampi}(2021)}]{chaithanya2021wall}%
  \BibitemOpen
  \bibfield  {author} {\bibinfo {author} {\bibfnamefont {K.}~\bibnamefont
  {Chaithanya}}\ and\ \bibinfo {author} {\bibfnamefont {S.~P.}\ \bibnamefont
  {Thampi}},\ }\bibfield  {title} {\bibinfo {title} {Wall-curvature driven
  dynamics of a microswimmer},\ }\href@noop {} {\bibfield  {journal} {\bibinfo
  {journal} {Physical Review Fluids}\ }\textbf {\bibinfo {volume} {6}},\
  \bibinfo {pages} {083101} (\bibinfo {year} {2021})}\BibitemShut {NoStop}%
\bibitem [{\citenamefont {Takaha}\ and\ \citenamefont
  {Nishiguchi}(2022)}]{Tahaka2022}%
  \BibitemOpen
  \bibfield  {author} {\bibinfo {author} {\bibfnamefont {Y.}~\bibnamefont
  {Takaha}}\ and\ \bibinfo {author} {\bibfnamefont {D.}~\bibnamefont
  {Nishiguchi}},\ }\bibfield  {title} {\bibinfo {title} {Quasi-two-dimensional
  bacterial swimming around pillars: enhanced trapping efficiency and curvature
  dependence},\ }\href@noop {} {\bibfield  {journal} {\bibinfo  {journal}
  {arXiv preprint arXiv:2203.16017}\ } (\bibinfo {year} {2022})}\BibitemShut
  {NoStop}%
\bibitem [{\citenamefont {Ketzetzi}\ \emph {et~al.}(2022)\citenamefont
  {Ketzetzi}, \citenamefont {Rinaldin}, \citenamefont {Dr{\"o}ge},
  \citenamefont {Graaf},\ and\ \citenamefont {Kraft}}]{ketzetzi2022activity}%
  \BibitemOpen
  \bibfield  {author} {\bibinfo {author} {\bibfnamefont {S.}~\bibnamefont
  {Ketzetzi}}, \bibinfo {author} {\bibfnamefont {M.}~\bibnamefont {Rinaldin}},
  \bibinfo {author} {\bibfnamefont {P.}~\bibnamefont {Dr{\"o}ge}}, \bibinfo
  {author} {\bibfnamefont {J.~d.}\ \bibnamefont {Graaf}},\ and\ \bibinfo
  {author} {\bibfnamefont {D.~J.}\ \bibnamefont {Kraft}},\ }\bibfield  {title}
  {\bibinfo {title} {Activity-induced interactions and cooperation of
  artificial microswimmers in one-dimensional environments},\ }\href@noop {}
  {\bibfield  {journal} {\bibinfo  {journal} {Nature Communications}\ }\textbf
  {\bibinfo {volume} {13}},\ \bibinfo {pages} {1} (\bibinfo {year}
  {2022})}\BibitemShut {NoStop}%
\bibitem [{\citenamefont {Fa{\'u}ndez}\ \emph {et~al.}(2022)\citenamefont
  {Fa{\'u}ndez}, \citenamefont {Espinoza}, \citenamefont {Soto},\ and\
  \citenamefont {Guzm{\'a}n-Lastra}}]{faundez2022microbial}%
  \BibitemOpen
  \bibfield  {author} {\bibinfo {author} {\bibfnamefont {T.}~\bibnamefont
  {Fa{\'u}ndez}}, \bibinfo {author} {\bibfnamefont {B.}~\bibnamefont
  {Espinoza}}, \bibinfo {author} {\bibfnamefont {R.}~\bibnamefont {Soto}},\
  and\ \bibinfo {author} {\bibfnamefont {F.}~\bibnamefont
  {Guzm{\'a}n-Lastra}},\ }\bibfield  {title} {\bibinfo {title} {Microbial
  adhesion on circular obstacles: An optimization study},\ }\href@noop {}
  {\bibfield  {journal} {\bibinfo  {journal} {Frontiers in Physics}\ }\textbf
  {\bibinfo {volume} {10}},\ \bibinfo {pages} {865937} (\bibinfo {year}
  {2022})}\BibitemShut {NoStop}%
\bibitem [{\citenamefont {Delmotte}\ \emph {et~al.}(2017)\citenamefont
  {Delmotte}, \citenamefont {Donev}, \citenamefont {Driscoll},\ and\
  \citenamefont {Chaikin}}]{delmotte2017minimal}%
  \BibitemOpen
  \bibfield  {author} {\bibinfo {author} {\bibfnamefont {B.}~\bibnamefont
  {Delmotte}}, \bibinfo {author} {\bibfnamefont {A.}~\bibnamefont {Donev}},
  \bibinfo {author} {\bibfnamefont {M.}~\bibnamefont {Driscoll}},\ and\
  \bibinfo {author} {\bibfnamefont {P.}~\bibnamefont {Chaikin}},\ }\bibfield
  {title} {\bibinfo {title} {Minimal model for a hydrodynamic fingering
  instability in microroller suspensions},\ }\href@noop {} {\bibfield
  {journal} {\bibinfo  {journal} {Physical Review Fluids}\ }\textbf {\bibinfo
  {volume} {2}},\ \bibinfo {pages} {114301} (\bibinfo {year}
  {2017})}\BibitemShut {NoStop}%
\bibitem [{\citenamefont {Liebchen}\ and\ \citenamefont
  {Mukhopadhyay}(2021)}]{liebchen2021interactions}%
  \BibitemOpen
  \bibfield  {author} {\bibinfo {author} {\bibfnamefont {B.}~\bibnamefont
  {Liebchen}}\ and\ \bibinfo {author} {\bibfnamefont {A.~K.}\ \bibnamefont
  {Mukhopadhyay}},\ }\bibfield  {title} {\bibinfo {title} {Interactions in
  active colloids},\ }\href@noop {} {\bibfield  {journal} {\bibinfo  {journal}
  {Journal of Physics: Condensed Matter}\ }\textbf {\bibinfo {volume} {34}},\
  \bibinfo {pages} {083002} (\bibinfo {year} {2021})}\BibitemShut {NoStop}%
\bibitem [{\citenamefont {Martinez-Pedrero}\ \emph {et~al.}(2018)\citenamefont
  {Martinez-Pedrero}, \citenamefont {Navarro-Argem{\'\i}}, \citenamefont
  {Ortiz-Ambriz}, \citenamefont {Pagonabarraga},\ and\ \citenamefont
  {Tierno}}]{martinez2018emergent}%
  \BibitemOpen
  \bibfield  {author} {\bibinfo {author} {\bibfnamefont {F.}~\bibnamefont
  {Martinez-Pedrero}}, \bibinfo {author} {\bibfnamefont {E.}~\bibnamefont
  {Navarro-Argem{\'\i}}}, \bibinfo {author} {\bibfnamefont {A.}~\bibnamefont
  {Ortiz-Ambriz}}, \bibinfo {author} {\bibfnamefont {I.}~\bibnamefont
  {Pagonabarraga}},\ and\ \bibinfo {author} {\bibfnamefont {P.}~\bibnamefont
  {Tierno}},\ }\bibfield  {title} {\bibinfo {title} {Emergent hydrodynamic
  bound states between magnetically powered micropropellers},\ }\href@noop {}
  {\bibfield  {journal} {\bibinfo  {journal} {Science advances}\ }\textbf
  {\bibinfo {volume} {4}},\ \bibinfo {pages} {eaap9379} (\bibinfo {year}
  {2018})}\BibitemShut {NoStop}%
\bibitem [{\citenamefont {Delmotte}(2019)}]{delmotte2019hydrodynamically}%
  \BibitemOpen
  \bibfield  {author} {\bibinfo {author} {\bibfnamefont {B.}~\bibnamefont
  {Delmotte}},\ }\bibfield  {title} {\bibinfo {title} {Hydrodynamically bound
  states of a pair of microrollers: A dynamical system insight},\ }\href@noop
  {} {\bibfield  {journal} {\bibinfo  {journal} {Physical Review Fluids}\
  }\textbf {\bibinfo {volume} {4}},\ \bibinfo {pages} {044302} (\bibinfo {year}
  {2019})}\BibitemShut {NoStop}%
\bibitem [{\citenamefont {Sprinkle}\ \emph {et~al.}(2017)\citenamefont
  {Sprinkle}, \citenamefont {Balboa~Usabiaga}, \citenamefont {Patankar},\ and\
  \citenamefont {Donev}}]{sprinkle2017large}%
  \BibitemOpen
  \bibfield  {author} {\bibinfo {author} {\bibfnamefont {B.}~\bibnamefont
  {Sprinkle}}, \bibinfo {author} {\bibfnamefont {F.}~\bibnamefont
  {Balboa~Usabiaga}}, \bibinfo {author} {\bibfnamefont {N.~A.}\ \bibnamefont
  {Patankar}},\ and\ \bibinfo {author} {\bibfnamefont {A.}~\bibnamefont
  {Donev}},\ }\bibfield  {title} {\bibinfo {title} {Large scale brownian
  dynamics of confined suspensions of rigid particles},\ }\href@noop {}
  {\bibfield  {journal} {\bibinfo  {journal} {The Journal of Chemical Physics}\
  }\textbf {\bibinfo {volume} {147}},\ \bibinfo {pages} {244103} (\bibinfo
  {year} {2017})}\BibitemShut {NoStop}%
\bibitem [{\citenamefont {Sprinkle}\ \emph {et~al.}(2020)\citenamefont
  {Sprinkle}, \citenamefont {van~der Wee}, \citenamefont {Luo}, \citenamefont
  {Driscoll},\ and\ \citenamefont {Donev}}]{sprinkle2020active}%
  \BibitemOpen
  \bibfield  {author} {\bibinfo {author} {\bibfnamefont {B.}~\bibnamefont
  {Sprinkle}}, \bibinfo {author} {\bibfnamefont {E.~B.}\ \bibnamefont {van~der
  Wee}}, \bibinfo {author} {\bibfnamefont {Y.}~\bibnamefont {Luo}}, \bibinfo
  {author} {\bibfnamefont {M.~M.}\ \bibnamefont {Driscoll}},\ and\ \bibinfo
  {author} {\bibfnamefont {A.}~\bibnamefont {Donev}},\ }\bibfield  {title}
  {\bibinfo {title} {Driven dynamics in dense suspensions of microrollers},\
  }\href@noop {} {\bibfield  {journal} {\bibinfo  {journal} {Soft Matter}\
  }\textbf {\bibinfo {volume} {16}},\ \bibinfo {pages} {7982} (\bibinfo {year}
  {2020})}\BibitemShut {NoStop}%
\bibitem [{\citenamefont {Junot}\ \emph {et~al.}(2021)\citenamefont {Junot},
  \citenamefont {Cebers},\ and\ \citenamefont {Tierno}}]{junot2021collective}%
  \BibitemOpen
  \bibfield  {author} {\bibinfo {author} {\bibfnamefont {G.}~\bibnamefont
  {Junot}}, \bibinfo {author} {\bibfnamefont {A.}~\bibnamefont {Cebers}},\ and\
  \bibinfo {author} {\bibfnamefont {P.}~\bibnamefont {Tierno}},\ }\bibfield
  {title} {\bibinfo {title} {Collective hydrodynamic transport of magnetic
  microrollers},\ }\href@noop {} {\bibfield  {journal} {\bibinfo  {journal}
  {Soft Matter}\ }\textbf {\bibinfo {volume} {17}},\ \bibinfo {pages} {8605}
  (\bibinfo {year} {2021})}\BibitemShut {NoStop}%
\bibitem [{\citenamefont {Alapan}\ \emph {et~al.}(2020)\citenamefont {Alapan},
  \citenamefont {Bozuyuk}, \citenamefont {Erkoc}, \citenamefont {Karacakol},\
  and\ \citenamefont {Sitti}}]{alapan2020multifunctional}%
  \BibitemOpen
  \bibfield  {author} {\bibinfo {author} {\bibfnamefont {Y.}~\bibnamefont
  {Alapan}}, \bibinfo {author} {\bibfnamefont {U.}~\bibnamefont {Bozuyuk}},
  \bibinfo {author} {\bibfnamefont {P.}~\bibnamefont {Erkoc}}, \bibinfo
  {author} {\bibfnamefont {A.~C.}\ \bibnamefont {Karacakol}},\ and\ \bibinfo
  {author} {\bibfnamefont {M.}~\bibnamefont {Sitti}},\ }\bibfield  {title}
  {\bibinfo {title} {Multifunctional surface microrollers for targeted cargo
  delivery in physiological blood flow},\ }\href@noop {} {\bibfield  {journal}
  {\bibinfo  {journal} {Science Robotics}\ }\textbf {\bibinfo {volume} {5}}
  (\bibinfo {year} {2020})}\BibitemShut {NoStop}%
\bibitem [{\citenamefont {Balboa~Usabiaga}\ \emph
  {et~al.}(2017{\natexlab{a}})\citenamefont {Balboa~Usabiaga}, \citenamefont
  {Kallemov}, \citenamefont {Delmotte}, \citenamefont {Bhalla}, \citenamefont
  {Griffith},\ and\ \citenamefont {Donev}}]{balboa2017hydrodynamics}%
  \BibitemOpen
  \bibfield  {author} {\bibinfo {author} {\bibfnamefont {F.}~\bibnamefont
  {Balboa~Usabiaga}}, \bibinfo {author} {\bibfnamefont {B.}~\bibnamefont
  {Kallemov}}, \bibinfo {author} {\bibfnamefont {B.}~\bibnamefont {Delmotte}},
  \bibinfo {author} {\bibfnamefont {A.}~\bibnamefont {Bhalla}}, \bibinfo
  {author} {\bibfnamefont {B.}~\bibnamefont {Griffith}},\ and\ \bibinfo
  {author} {\bibfnamefont {A.}~\bibnamefont {Donev}},\ }\bibfield  {title}
  {\bibinfo {title} {Hydrodynamics of suspensions of passive and active rigid
  particles: a rigid multiblob approach},\ }\href@noop {} {\bibfield  {journal}
  {\bibinfo  {journal} {Communications in Applied Mathematics and Computational
  Science}\ }\textbf {\bibinfo {volume} {11}},\ \bibinfo {pages} {217}
  (\bibinfo {year} {2017}{\natexlab{a}})}\BibitemShut {NoStop}%
\bibitem [{\citenamefont {Chamolly}\ \emph {et~al.}(2020)\citenamefont
  {Chamolly}, \citenamefont {Lauga},\ and\ \citenamefont
  {Tottori}}]{chamolly2020irreversible}%
  \BibitemOpen
  \bibfield  {author} {\bibinfo {author} {\bibfnamefont {A.}~\bibnamefont
  {Chamolly}}, \bibinfo {author} {\bibfnamefont {E.}~\bibnamefont {Lauga}},\
  and\ \bibinfo {author} {\bibfnamefont {S.}~\bibnamefont {Tottori}},\
  }\bibfield  {title} {\bibinfo {title} {Irreversible hydrodynamic trapping by
  surface rollers},\ }\href@noop {} {\bibfield  {journal} {\bibinfo  {journal}
  {Soft matter}\ }\textbf {\bibinfo {volume} {16}},\ \bibinfo {pages} {2611}
  (\bibinfo {year} {2020})}\BibitemShut {NoStop}%
\bibitem [{\citenamefont {Demir{\"o}rs}\ \emph {et~al.}(2021)\citenamefont
  {Demir{\"o}rs}, \citenamefont {Stauffer}, \citenamefont {Lauener},
  \citenamefont {Cossu}, \citenamefont {Ramakrishna}, \citenamefont {de~Graaf},
  \citenamefont {Alcantara}, \citenamefont {Pan{\'e}}, \citenamefont
  {Spencer},\ and\ \citenamefont {Studart}}]{demirors2021magnetic}%
  \BibitemOpen
  \bibfield  {author} {\bibinfo {author} {\bibfnamefont {A.~F.}\ \bibnamefont
  {Demir{\"o}rs}}, \bibinfo {author} {\bibfnamefont {A.}~\bibnamefont
  {Stauffer}}, \bibinfo {author} {\bibfnamefont {C.}~\bibnamefont {Lauener}},
  \bibinfo {author} {\bibfnamefont {J.}~\bibnamefont {Cossu}}, \bibinfo
  {author} {\bibfnamefont {S.~N.}\ \bibnamefont {Ramakrishna}}, \bibinfo
  {author} {\bibfnamefont {J.}~\bibnamefont {de~Graaf}}, \bibinfo {author}
  {\bibfnamefont {C.~C.}\ \bibnamefont {Alcantara}}, \bibinfo {author}
  {\bibfnamefont {S.}~\bibnamefont {Pan{\'e}}}, \bibinfo {author}
  {\bibfnamefont {N.}~\bibnamefont {Spencer}},\ and\ \bibinfo {author}
  {\bibfnamefont {A.~R.}\ \bibnamefont {Studart}},\ }\bibfield  {title}
  {\bibinfo {title} {Magnetic propulsion of colloidal microrollers controlled
  by electrically modulated friction},\ }\href@noop {} {\bibfield  {journal}
  {\bibinfo  {journal} {Soft Matter}\ }\textbf {\bibinfo {volume} {17}},\
  \bibinfo {pages} {1037} (\bibinfo {year} {2021})}\BibitemShut {NoStop}%
\bibitem [{\citenamefont {Bozuyuk}\ \emph
  {et~al.}(2022{\natexlab{a}})\citenamefont {Bozuyuk}, \citenamefont {Suadiye},
  \citenamefont {Aghakhani}, \citenamefont {Dogan}, \citenamefont {Lazovic},
  \citenamefont {Tiryaki}, \citenamefont {Schneider}, \citenamefont
  {Karacakol}, \citenamefont {Demir}, \citenamefont {Richter} \emph
  {et~al.}}]{bozuyuk2022high}%
  \BibitemOpen
  \bibfield  {author} {\bibinfo {author} {\bibfnamefont {U.}~\bibnamefont
  {Bozuyuk}}, \bibinfo {author} {\bibfnamefont {E.}~\bibnamefont {Suadiye}},
  \bibinfo {author} {\bibfnamefont {A.}~\bibnamefont {Aghakhani}}, \bibinfo
  {author} {\bibfnamefont {N.~O.}\ \bibnamefont {Dogan}}, \bibinfo {author}
  {\bibfnamefont {J.}~\bibnamefont {Lazovic}}, \bibinfo {author} {\bibfnamefont
  {M.~E.}\ \bibnamefont {Tiryaki}}, \bibinfo {author} {\bibfnamefont
  {M.}~\bibnamefont {Schneider}}, \bibinfo {author} {\bibfnamefont {A.~C.}\
  \bibnamefont {Karacakol}}, \bibinfo {author} {\bibfnamefont {S.~O.}\
  \bibnamefont {Demir}}, \bibinfo {author} {\bibfnamefont {G.}~\bibnamefont
  {Richter}}, \emph {et~al.},\ }\bibfield  {title} {\bibinfo {title}
  {High-performance magnetic fept (l10) surface microrollers towards medical
  imaging-guided endovascular delivery applications},\ }\href@noop {}
  {\bibfield  {journal} {\bibinfo  {journal} {Advanced Functional Materials}\
  }\textbf {\bibinfo {volume} {32}},\ \bibinfo {pages} {2109741} (\bibinfo
  {year} {2022}{\natexlab{a}})}\BibitemShut {NoStop}%
\bibitem [{\citenamefont {Bozuyuk}\ \emph
  {et~al.}(2022{\natexlab{b}})\citenamefont {Bozuyuk}, \citenamefont
  {Aghakhani}, \citenamefont {Alapan}, \citenamefont {Yunusa}, \citenamefont
  {Wrede},\ and\ \citenamefont {Sitti}}]{bozuyuk2022reduced}%
  \BibitemOpen
  \bibfield  {author} {\bibinfo {author} {\bibfnamefont {U.}~\bibnamefont
  {Bozuyuk}}, \bibinfo {author} {\bibfnamefont {A.}~\bibnamefont {Aghakhani}},
  \bibinfo {author} {\bibfnamefont {Y.}~\bibnamefont {Alapan}}, \bibinfo
  {author} {\bibfnamefont {M.}~\bibnamefont {Yunusa}}, \bibinfo {author}
  {\bibfnamefont {P.}~\bibnamefont {Wrede}},\ and\ \bibinfo {author}
  {\bibfnamefont {M.}~\bibnamefont {Sitti}},\ }\bibfield  {title} {\bibinfo
  {title} {Reduced rotational flows enable the translation of surface-rolling
  microrobots in confined spaces},\ }\href@noop {} {\bibfield  {journal}
  {\bibinfo  {journal} {Nature communications}\ }\textbf {\bibinfo {volume}
  {13}},\ \bibinfo {pages} {1} (\bibinfo {year}
  {2022}{\natexlab{b}})}\BibitemShut {NoStop}%
\bibitem [{\citenamefont {Balboa~Usabiaga}\ \emph
  {et~al.}(2017{\natexlab{b}})\citenamefont {Balboa~Usabiaga}, \citenamefont
  {Delmotte},\ and\ \citenamefont {Donev}}]{balboa2017brownian}%
  \BibitemOpen
  \bibfield  {author} {\bibinfo {author} {\bibfnamefont {F.}~\bibnamefont
  {Balboa~Usabiaga}}, \bibinfo {author} {\bibfnamefont {B.}~\bibnamefont
  {Delmotte}},\ and\ \bibinfo {author} {\bibfnamefont {A.}~\bibnamefont
  {Donev}},\ }\bibfield  {title} {\bibinfo {title} {Brownian dynamics of
  confined suspensions of active microrollers},\ }\href@noop {} {\bibfield
  {journal} {\bibinfo  {journal} {The Journal of Chemical Physics}\ }\textbf
  {\bibinfo {volume} {146}},\ \bibinfo {pages} {134104} (\bibinfo {year}
  {2017}{\natexlab{b}})}\BibitemShut {NoStop}%
\bibitem [{Note1()}]{Note1}%
  \BibitemOpen
  \bibinfo {note} {See Supplemental Files at \protect \url
  {https://doi.org/10.6084/m9.figshare.19772950}}\BibitemShut {NoStop}%
\bibitem [{\citenamefont {van~der Wel}\ \emph {et~al.}(2017)\citenamefont
  {van~der Wel}, \citenamefont {Bhan}, \citenamefont {Verweij}, \citenamefont
  {Frijters}, \citenamefont {Gong}, \citenamefont {Hollingsworth},
  \citenamefont {Sacanna},\ and\ \citenamefont {Kraft}}]{van2017preparation}%
  \BibitemOpen
  \bibfield  {author} {\bibinfo {author} {\bibfnamefont {C.}~\bibnamefont
  {van~der Wel}}, \bibinfo {author} {\bibfnamefont {R.~K.}\ \bibnamefont
  {Bhan}}, \bibinfo {author} {\bibfnamefont {R.~W.}\ \bibnamefont {Verweij}},
  \bibinfo {author} {\bibfnamefont {H.~C.}\ \bibnamefont {Frijters}}, \bibinfo
  {author} {\bibfnamefont {Z.}~\bibnamefont {Gong}}, \bibinfo {author}
  {\bibfnamefont {A.~D.}\ \bibnamefont {Hollingsworth}}, \bibinfo {author}
  {\bibfnamefont {S.}~\bibnamefont {Sacanna}},\ and\ \bibinfo {author}
  {\bibfnamefont {D.~J.}\ \bibnamefont {Kraft}},\ }\bibfield  {title} {\bibinfo
  {title} {Preparation of colloidal organosilica spheres through spontaneous
  emulsification},\ }\href@noop {} {\bibfield  {journal} {\bibinfo  {journal}
  {Langmuir}\ }\textbf {\bibinfo {volume} {33}},\ \bibinfo {pages} {8174}
  (\bibinfo {year} {2017})}\BibitemShut {NoStop}%
\bibitem [{\citenamefont {Baker}\ \emph {et~al.}(2019)\citenamefont {Baker},
  \citenamefont {Montenegro-Johnson}, \citenamefont {Sediako}, \citenamefont
  {Thomson}, \citenamefont {Sen}, \citenamefont {Lauga},\ and\ \citenamefont
  {Aranson}}]{baker2019shape}%
  \BibitemOpen
  \bibfield  {author} {\bibinfo {author} {\bibfnamefont {R.~D.}\ \bibnamefont
  {Baker}}, \bibinfo {author} {\bibfnamefont {T.}~\bibnamefont
  {Montenegro-Johnson}}, \bibinfo {author} {\bibfnamefont {A.~D.}\ \bibnamefont
  {Sediako}}, \bibinfo {author} {\bibfnamefont {M.~J.}\ \bibnamefont
  {Thomson}}, \bibinfo {author} {\bibfnamefont {A.}~\bibnamefont {Sen}},
  \bibinfo {author} {\bibfnamefont {E.}~\bibnamefont {Lauga}},\ and\ \bibinfo
  {author} {\bibfnamefont {I.~S.}\ \bibnamefont {Aranson}},\ }\bibfield
  {title} {\bibinfo {title} {Shape-programmed 3d printed swimming microtori for
  the transport of passive and active agents},\ }\href@noop {} {\bibfield
  {journal} {\bibinfo  {journal} {Nature communications}\ }\textbf {\bibinfo
  {volume} {10}},\ \bibinfo {pages} {1} (\bibinfo {year} {2019})}\BibitemShut
  {NoStop}%
\bibitem [{\citenamefont {Delong}\ \emph {et~al.}(2020)\citenamefont {Delong},
  \citenamefont {Balboa~Usabiaga}, \citenamefont {Delmotte}, \citenamefont
  {Sprinkle},\ and\ \citenamefont {Donev}}]{RigidMultiblobs}%
  \BibitemOpen
  \bibfield  {author} {\bibinfo {author} {\bibfnamefont {S.}~\bibnamefont
  {Delong}}, \bibinfo {author} {\bibfnamefont {F.}~\bibnamefont
  {Balboa~Usabiaga}}, \bibinfo {author} {\bibfnamefont {B.}~\bibnamefont
  {Delmotte}}, \bibinfo {author} {\bibfnamefont {B.}~\bibnamefont {Sprinkle}},\
  and\ \bibinfo {author} {\bibfnamefont {A.}~\bibnamefont {Donev}},\
  }\href@noop {} {\bibinfo {title} {Rigid multiblobs in half-space}},\ \bibinfo
  {howpublished} {\url{
  https://github.com/stochasticHydroTools/RigidMultiblobsWall}} (\bibinfo
  {year} {2020})\BibitemShut {NoStop}%
\bibitem [{geo()}]{geometric}%
  \BibitemOpen
  \href@noop {} {}\bibinfo {note} {Equivalent to a 0.792 \textmu m geometric
  radius.}\BibitemShut {Stop}%
\bibitem [{\citenamefont {Crocker}\ and\ \citenamefont
  {Grier}(1996)}]{crocker1996methods}%
  \BibitemOpen
  \bibfield  {author} {\bibinfo {author} {\bibfnamefont {J.~C.}\ \bibnamefont
  {Crocker}}\ and\ \bibinfo {author} {\bibfnamefont {D.~G.}\ \bibnamefont
  {Grier}},\ }\bibfield  {title} {\bibinfo {title} {Methods of digital video
  microscopy for colloidal studies},\ }\href@noop {} {\bibfield  {journal}
  {\bibinfo  {journal} {Journal of colloid and interface science}\ }\textbf
  {\bibinfo {volume} {179}},\ \bibinfo {pages} {298} (\bibinfo {year}
  {1996})}\BibitemShut {NoStop}%
\bibitem [{\citenamefont {Allan}\ \emph {et~al.}(2018)\citenamefont {Allan},
  \citenamefont {Caswell}, \citenamefont {Keim},\ and\ \citenamefont {van~der
  Wel}}]{allan2014trackpy}%
  \BibitemOpen
  \bibfield  {author} {\bibinfo {author} {\bibfnamefont {D.}~\bibnamefont
  {Allan}}, \bibinfo {author} {\bibfnamefont {T.}~\bibnamefont {Caswell}},
  \bibinfo {author} {\bibfnamefont {N.}~\bibnamefont {Keim}},\ and\ \bibinfo
  {author} {\bibfnamefont {C.}~\bibnamefont {van~der Wel}},\ }\bibfield
  {title} {\bibinfo {title} {Trackpy},\ }\href@noop {} {\bibfield  {journal}
  {\bibinfo  {journal} {DOI: \url{https://doi.org/10.5281/zenodo.1226458}}\ }
  (\bibinfo {year} {2018})}\BibitemShut {NoStop}%
\bibitem [{esc()}]{escape}%
  \BibitemOpen
  \href@noop {} {}\bibinfo {note} {As a passing roller does not always enter
  the hydrodynamic trap, we measure the escape time by placing the roller
  inside the trap to save the run-time of the simulations.}\BibitemShut {Stop}%
\bibitem [{cri()}]{criterion}%
  \BibitemOpen
  \href@noop {} {}\bibinfo {note} {We did not find the escape time to be very
  sensitive to the distance used define escape. Using a smaller escape
  criterion $x > R_h + 4r$ we found a change in the measured escape times of
  $-5 \pm 8 \%$.}\BibitemShut {Stop}%
\bibitem [{fre()}]{freepassage}%
  \BibitemOpen
  \href@noop {} {}\bibinfo {note} {The escape time without an obstacle present
  is the mean time a free roller takes to translate five times its own radius.
  This corresponds to $\langle t_{esc}^{no~obstacle}\rangle = (R_h+5r_h)/V_0$,
  where $V_0$ is the velocity of the roller in absence of an
  obstacle.}\BibitemShut {Stop}%
\bibitem [{\citenamefont {Yethiraj}\ and\ \citenamefont {van
  Blaaderen}(2003)}]{yethiraj2003colloidal}%
  \BibitemOpen
  \bibfield  {author} {\bibinfo {author} {\bibfnamefont {A.}~\bibnamefont
  {Yethiraj}}\ and\ \bibinfo {author} {\bibfnamefont {A.}~\bibnamefont {van
  Blaaderen}},\ }\bibfield  {title} {\bibinfo {title} {A colloidal model system
  with an interaction tunable from hard sphere to soft and dipolar},\
  }\href@noop {} {\bibfield  {journal} {\bibinfo  {journal} {nature}\ }\textbf
  {\bibinfo {volume} {421}},\ \bibinfo {pages} {513} (\bibinfo {year}
  {2003})}\BibitemShut {NoStop}%
\bibitem [{ele()}]{electro}%
  \BibitemOpen
  \href@noop {} {}\bibinfo {note} {The value of $d=0.8r_h$ was chosen such that
  microrollers at positions $x^2+y^2>(R_h+r_h+d)^2$ did not have a velocity
  larger than their self-induced velocity in bulk $V_0$.}\BibitemShut {Stop}%
\bibitem [{\citenamefont {Grier}(1997)}]{grier1997optical}%
  \BibitemOpen
  \bibfield  {author} {\bibinfo {author} {\bibfnamefont {D.~G.}\ \bibnamefont
  {Grier}},\ }\bibfield  {title} {\bibinfo {title} {Optical tweezers in colloid
  and interface science},\ }\href@noop {} {\bibfield  {journal} {\bibinfo
  {journal} {Current opinion in colloid \& interface science}\ }\textbf
  {\bibinfo {volume} {2}},\ \bibinfo {pages} {264} (\bibinfo {year}
  {1997})}\BibitemShut {NoStop}%
\bibitem [{\citenamefont {Saraswat}\ \emph {et~al.}(2020)\citenamefont
  {Saraswat}, \citenamefont {Ibis}, \citenamefont {Rossi}, \citenamefont
  {Sasso}, \citenamefont {Eral},\ and\ \citenamefont
  {Fanzio}}]{saraswat2020shape}%
  \BibitemOpen
  \bibfield  {author} {\bibinfo {author} {\bibfnamefont {Y.~C.}\ \bibnamefont
  {Saraswat}}, \bibinfo {author} {\bibfnamefont {F.}~\bibnamefont {Ibis}},
  \bibinfo {author} {\bibfnamefont {L.}~\bibnamefont {Rossi}}, \bibinfo
  {author} {\bibfnamefont {L.}~\bibnamefont {Sasso}}, \bibinfo {author}
  {\bibfnamefont {H.~B.}\ \bibnamefont {Eral}},\ and\ \bibinfo {author}
  {\bibfnamefont {P.}~\bibnamefont {Fanzio}},\ }\bibfield  {title} {\bibinfo
  {title} {Shape anisotropic colloidal particle fabrication using 2-photon
  polymerization},\ }\href@noop {} {\bibfield  {journal} {\bibinfo  {journal}
  {Journal of colloid and interface science}\ }\textbf {\bibinfo {volume}
  {564}},\ \bibinfo {pages} {43} (\bibinfo {year} {2020})}\BibitemShut
  {NoStop}%
\bibitem [{\citenamefont {Doherty}\ \emph {et~al.}(2020)\citenamefont
  {Doherty}, \citenamefont {Varkevisser}, \citenamefont {Teunisse},
  \citenamefont {Hoecht}, \citenamefont {Ketzetzi}, \citenamefont {Ouhajji},\
  and\ \citenamefont {Kraft}}]{doherty2020catalytically}%
  \BibitemOpen
  \bibfield  {author} {\bibinfo {author} {\bibfnamefont {R.~P.}\ \bibnamefont
  {Doherty}}, \bibinfo {author} {\bibfnamefont {T.}~\bibnamefont
  {Varkevisser}}, \bibinfo {author} {\bibfnamefont {M.}~\bibnamefont
  {Teunisse}}, \bibinfo {author} {\bibfnamefont {J.}~\bibnamefont {Hoecht}},
  \bibinfo {author} {\bibfnamefont {S.}~\bibnamefont {Ketzetzi}}, \bibinfo
  {author} {\bibfnamefont {S.}~\bibnamefont {Ouhajji}},\ and\ \bibinfo {author}
  {\bibfnamefont {D.~J.}\ \bibnamefont {Kraft}},\ }\bibfield  {title} {\bibinfo
  {title} {Catalytically propelled 3d printed colloidal microswimmers},\
  }\href@noop {} {\bibfield  {journal} {\bibinfo  {journal} {Soft Matter}\
  }\textbf {\bibinfo {volume} {16}},\ \bibinfo {pages} {10463} (\bibinfo {year}
  {2020})}\BibitemShut {NoStop}%
\bibitem [{\citenamefont {Sharifi-Mood}\ \emph {et~al.}(2017)\citenamefont
  {Sharifi-Mood}, \citenamefont {D{\'\i}az-Hyland},\ and\ \citenamefont
  {C{\'o}rdova-Figueroa}}]{sharifi2017dynamics}%
  \BibitemOpen
  \bibfield  {author} {\bibinfo {author} {\bibfnamefont {N.}~\bibnamefont
  {Sharifi-Mood}}, \bibinfo {author} {\bibfnamefont {P.~G.}\ \bibnamefont
  {D{\'\i}az-Hyland}},\ and\ \bibinfo {author} {\bibfnamefont {U.~M.}\
  \bibnamefont {C{\'o}rdova-Figueroa}},\ }\bibfield  {title} {\bibinfo {title}
  {Dynamics of a microswimmer near a curved wall: guided and trapped
  locomotions},\ }\href@noop {} {\bibfield  {journal} {\bibinfo  {journal}
  {arXiv preprint arXiv:1710.10578}\ } (\bibinfo {year} {2017})}\BibitemShut
  {NoStop}%
\bibitem [{TPM()}]{TPM}%
  \BibitemOpen
  \href@noop {} {}\bibinfo {note} {The TPM spheres had a total diameter of
  2.1$\pm$0.1 \textmu m and the hematite cubes a side length of 0.77 $\pm$ 0.1
  \textmu m, both measured using scanning electron microscopy (SEM). The TPM
  spheres where fluorescently labeled for fluorescence microscopy using
  4-methylaminoethylmethacrylate-7-nitrobenzo-2-oxa-1,3-diazol
  (NBDMAEM)~\cite{sprinkle2020active}.}\BibitemShut {Stop}%
\bibitem [{\citenamefont {Nishiguchi}\ \emph {et~al.}(2018)\citenamefont
  {Nishiguchi}, \citenamefont {Aranson}, \citenamefont {Snezhko},\ and\
  \citenamefont {Sokolov}}]{nishiguchi2018engineering}%
  \BibitemOpen
  \bibfield  {author} {\bibinfo {author} {\bibfnamefont {D.}~\bibnamefont
  {Nishiguchi}}, \bibinfo {author} {\bibfnamefont {I.~S.}\ \bibnamefont
  {Aranson}}, \bibinfo {author} {\bibfnamefont {A.}~\bibnamefont {Snezhko}},\
  and\ \bibinfo {author} {\bibfnamefont {A.}~\bibnamefont {Sokolov}},\
  }\bibfield  {title} {\bibinfo {title} {Engineering bacterial vortex lattice
  via direct laser lithography},\ }\href@noop {} {\bibfield  {journal}
  {\bibinfo  {journal} {Nature communications}\ }\textbf {\bibinfo {volume}
  {9}},\ \bibinfo {pages} {1} (\bibinfo {year} {2018})}\BibitemShut {NoStop}%
\bibitem [{\citenamefont {Reinken}\ \emph {et~al.}(2020)\citenamefont
  {Reinken}, \citenamefont {Nishiguchi}, \citenamefont {Heidenreich},
  \citenamefont {Sokolov}, \citenamefont {B{\"a}r}, \citenamefont {Klapp},\
  and\ \citenamefont {Aranson}}]{reinken2020organizing}%
  \BibitemOpen
  \bibfield  {author} {\bibinfo {author} {\bibfnamefont {H.}~\bibnamefont
  {Reinken}}, \bibinfo {author} {\bibfnamefont {D.}~\bibnamefont {Nishiguchi}},
  \bibinfo {author} {\bibfnamefont {S.}~\bibnamefont {Heidenreich}}, \bibinfo
  {author} {\bibfnamefont {A.}~\bibnamefont {Sokolov}}, \bibinfo {author}
  {\bibfnamefont {M.}~\bibnamefont {B{\"a}r}}, \bibinfo {author} {\bibfnamefont
  {S.~H.}\ \bibnamefont {Klapp}},\ and\ \bibinfo {author} {\bibfnamefont
  {I.~S.}\ \bibnamefont {Aranson}},\ }\bibfield  {title} {\bibinfo {title}
  {Organizing bacterial vortex lattices by periodic obstacle arrays},\
  }\href@noop {} {\bibfield  {journal} {\bibinfo  {journal} {Communications
  Physics}\ }\textbf {\bibinfo {volume} {3}},\ \bibinfo {pages} {1} (\bibinfo
  {year} {2020})}\BibitemShut {NoStop}%
\bibitem [{\citenamefont {van~der Walt}\ \emph {et~al.}(2014)\citenamefont
  {van~der Walt}, \citenamefont {{S}ch\"onberger}, \citenamefont
  {{Nunez-Iglesias}}, \citenamefont {{B}oulogne}, \citenamefont {{W}arner},
  \citenamefont {{Y}ager}, \citenamefont {{G}ouillart}, \citenamefont {{Y}u},\
  and\ \citenamefont {the scikit-image contributors}}]{scikit-image}%
  \BibitemOpen
  \bibfield  {author} {\bibinfo {author} {\bibfnamefont {S.}~\bibnamefont
  {van~der Walt}}, \bibinfo {author} {\bibfnamefont {J.~L.}\ \bibnamefont
  {{S}ch\"onberger}}, \bibinfo {author} {\bibfnamefont {J.}~\bibnamefont
  {{Nunez-Iglesias}}}, \bibinfo {author} {\bibfnamefont {F.}~\bibnamefont
  {{B}oulogne}}, \bibinfo {author} {\bibfnamefont {J.~D.}\ \bibnamefont
  {{W}arner}}, \bibinfo {author} {\bibfnamefont {N.}~\bibnamefont {{Y}ager}},
  \bibinfo {author} {\bibfnamefont {E.}~\bibnamefont {{G}ouillart}}, \bibinfo
  {author} {\bibfnamefont {T.}~\bibnamefont {{Y}u}},\ and\ \bibinfo {author}
  {\bibnamefont {the scikit-image contributors}},\ }\bibfield  {title}
  {\bibinfo {title} {scikit-image: image processing in {P}ython},\ }\href
  {https://doi.org/10.7717/peerj.453} {\bibfield  {journal} {\bibinfo
  {journal} {PeerJ}\ }\textbf {\bibinfo {volume} {2}},\ \bibinfo {pages} {e453}
  (\bibinfo {year} {2014})}\BibitemShut {NoStop}%
\bibitem [{\citenamefont {Grassia}\ and\ \citenamefont
  {Hinch}(1996)}]{Grassia1996}%
  \BibitemOpen
  \bibfield  {author} {\bibinfo {author} {\bibfnamefont {P.}~\bibnamefont
  {Grassia}}\ and\ \bibinfo {author} {\bibfnamefont {E.~J.}\ \bibnamefont
  {Hinch}},\ }\bibfield  {title} {\bibinfo {title} {Computer simulations of
  polymer chain relaxation via {Brownian} motion},\ }\href@noop {} {\bibfield
  {journal} {\bibinfo  {journal} {Journal of Fluid Mechanics}\ }\textbf
  {\bibinfo {volume} {308}},\ \bibinfo {pages} {255} (\bibinfo {year}
  {1996})}\BibitemShut {NoStop}%
\bibitem [{\citenamefont {Swan}\ and\ \citenamefont {Brady}(2007)}]{Swan2007}%
  \BibitemOpen
  \bibfield  {author} {\bibinfo {author} {\bibfnamefont {J.~W.}\ \bibnamefont
  {Swan}}\ and\ \bibinfo {author} {\bibfnamefont {J.~F.}\ \bibnamefont
  {Brady}},\ }\bibfield  {title} {\bibinfo {title} {Simulation of
  hydrodynamically interacting particles near a no-slip boundary},\ }\href@noop
  {} {\bibfield  {journal} {\bibinfo  {journal} {Physics of Fluids
  (1994-present)}\ }\textbf {\bibinfo {volume} {19}},\ \bibinfo {pages}
  {113306} (\bibinfo {year} {2007})}\BibitemShut {NoStop}%
\bibitem [{Note2()}]{Note2}%
  \BibitemOpen
  \bibinfo {note} {Visualizations throughout this manuscript of the roller and
  obstacle in simulations were made using OVITO~\cite {ovito}}\BibitemShut
  {NoStop}%
\bibitem [{\citenamefont {Goodwin}(2009)}]{goodwin2009colloids}%
  \BibitemOpen
  \bibfield  {author} {\bibinfo {author} {\bibfnamefont {J.}~\bibnamefont
  {Goodwin}},\ }\href@noop {} {\emph {\bibinfo {title} {Colloids and interfaces
  with surfactants and polymers}}}\ (\bibinfo  {publisher} {John Wiley \&
  Sons},\ \bibinfo {year} {2009})\BibitemShut {NoStop}%
\bibitem [{\citenamefont {Stukowski}(2010)}]{ovito}%
  \BibitemOpen
  \bibfield  {author} {\bibinfo {author} {\bibfnamefont {A.}~\bibnamefont
  {Stukowski}},\ }\bibfield  {title} {\bibinfo {title} {{Visualization and
  analysis of atomistic simulation data with OVITO-the Open Visualization
  Tool}},\ }\bibfield  {journal} {\bibinfo  {journal} {{MODELLING AND
  SIMULATION IN MATERIALS SCIENCE AND ENGINEERING}}\ }\textbf {\bibinfo
  {volume} {{18}}},\ \href {https://doi.org/{10.1088/0965-0393/18/1/015012}}
  {{10.1088/0965-0393/18/1/015012}} (\bibinfo {year} {{2010}})\BibitemShut
  {NoStop}%
\end{thebibliography}%

\end{document}


\baselineskip24pt
\maketitle 

\clearpage

\section*{Supporting File legends}

\subsection*{Video S1}
\noindent A microroller getting trapped by a cylindrical obstacle as imaged by fluorescence microscopy. Eventually the roller is able to escape the trap. The arrow denotes the direction of propagation of the roller. The obstacle radius is 14.4 microns.

\subsection*{Video S2}
\noindent A microroller passing a cylindrical obstacle without getting trapped, as imaged by fluorescence microscopy. The arrow denotes the direction of propagation of the roller. The obstacle radius is 14.4 microns.

\subsection*{Video S3}
\noindent A microroller getting trapped by a cylindrical obstacle in simulations. Eventually the roller escapes the trap. The arrow denotes the direction of propagation of the roller. The obstacle's hydrodynamic radius is 10 microns.

\subsection*{Video S4}
\noindent A microroller passing a cylindrical obstacle without getting trapped in simulations. The arrow denotes the direction of propagation of the roller. The obstacle's hydrodynamic radius is 10 microns.

\subsection*{Video S5}
\noindent An escape time measurement: the roller is placed in the wake of the obstacle (i.e. in the trap) and the simulation runs until the roller escapes. 

\subsection*{File S1}
\noindent 3D plot of the trajectories of a roller with different initial positions interacting with an obstacle. The colors denote different initial heights. The majority of initial positions lead to the convergence of the roller into a single point (black sphere). 

\clearpage

\section*{Supporting Tables}

\begin{table*}[h]
\centering
\begin{tabular}{| l | c| c| }
\hline
Parameter & Value & Units\\
\hline
Blob radius $r_b$ (N=12) & 0.416 & \textmu m \\ \hline
Blob radius $r_b$ (N=42) & 0.244 & \textmu m \\ \hline
Roller geometric radius $r_g$ & 0.7921 & \textmu m \\ \hline
Roller hydrodynamic radius $r_h$ &  1.0 & \textmu m \\ \hline
Buoyant force on roller $mg$ & 0.0372 & pN \\ \hline
Viscosity $\eta$ & $0.96 \times 10^{-3} $ & \si{Pa \cdot s}  \\  \hline
Temperature $T$ & 22 & \textdegree C    \\\hline
Torque $\tau ^*$ & $1.36 \times 10^{-18}$ & \si{N \cdot m} \\\hline
Solver tolerance & $10^{-6}$ & \\\hline
Potential at contact $\epsilon$ & 0.03 & \si{pN~\cdot}~\textmu m \\\hline
Debye length $b$ & 0.1 & \textmu m \\
\hline
\end{tabular}
\caption{\textbf{Parameters used in simulations.}
$^*$Equivalent to a driving frequency of 9 Hz in the absence of boundaries.
}
\end{table*}

\clearpage

\section*{Supporting Figures}


\begin{figure*}[h]
\centering
\includegraphics[width=\textwidth]{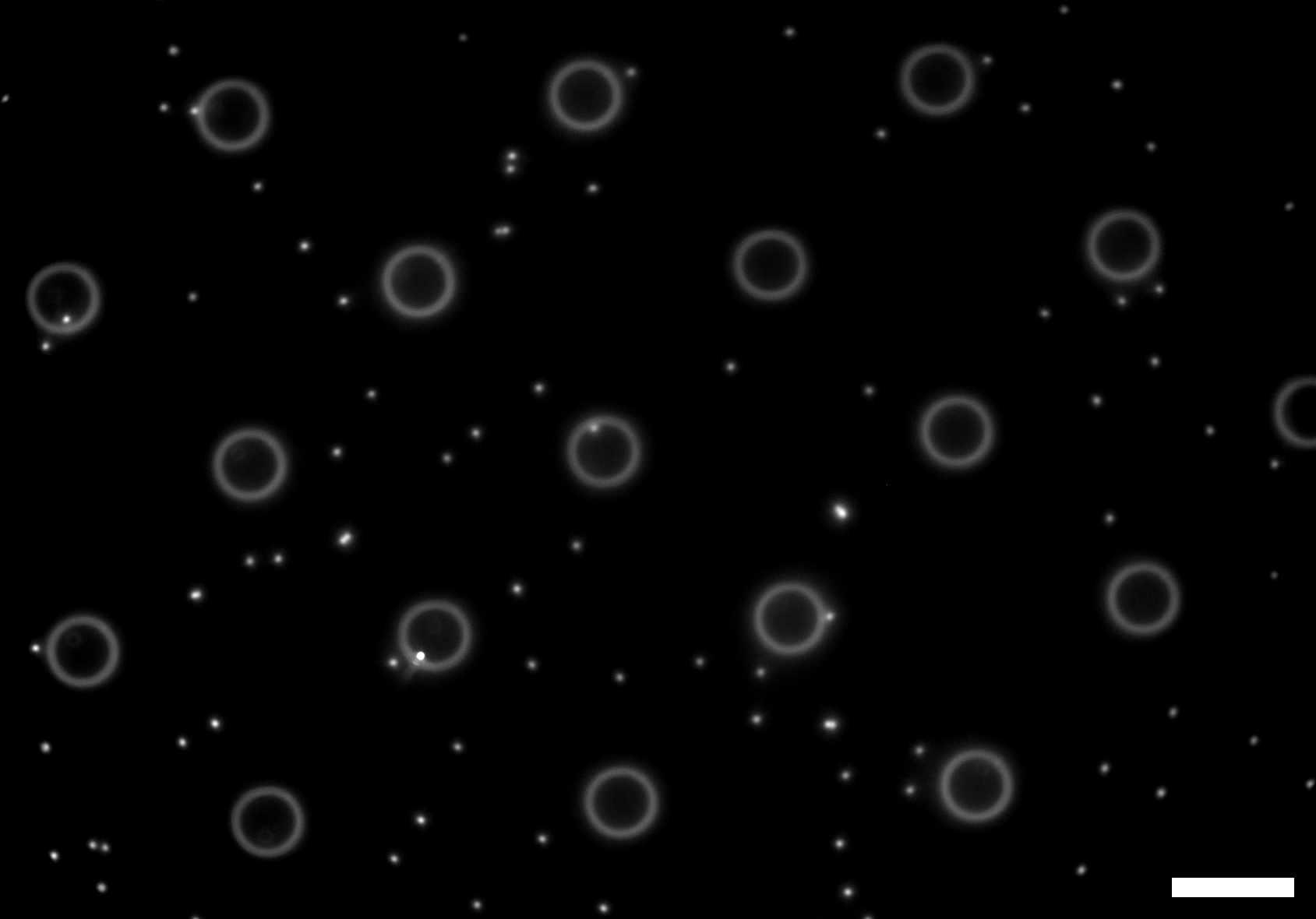}
\caption{\label{fig:SI_pillars}
\textbf{Fluorescence microscopy image of an array of printed pillars and microrollers.} The scale bar is 50 \textmu m.
}
\end{figure*}

\begin{figure*}
\centering
\includegraphics[width=\textwidth]{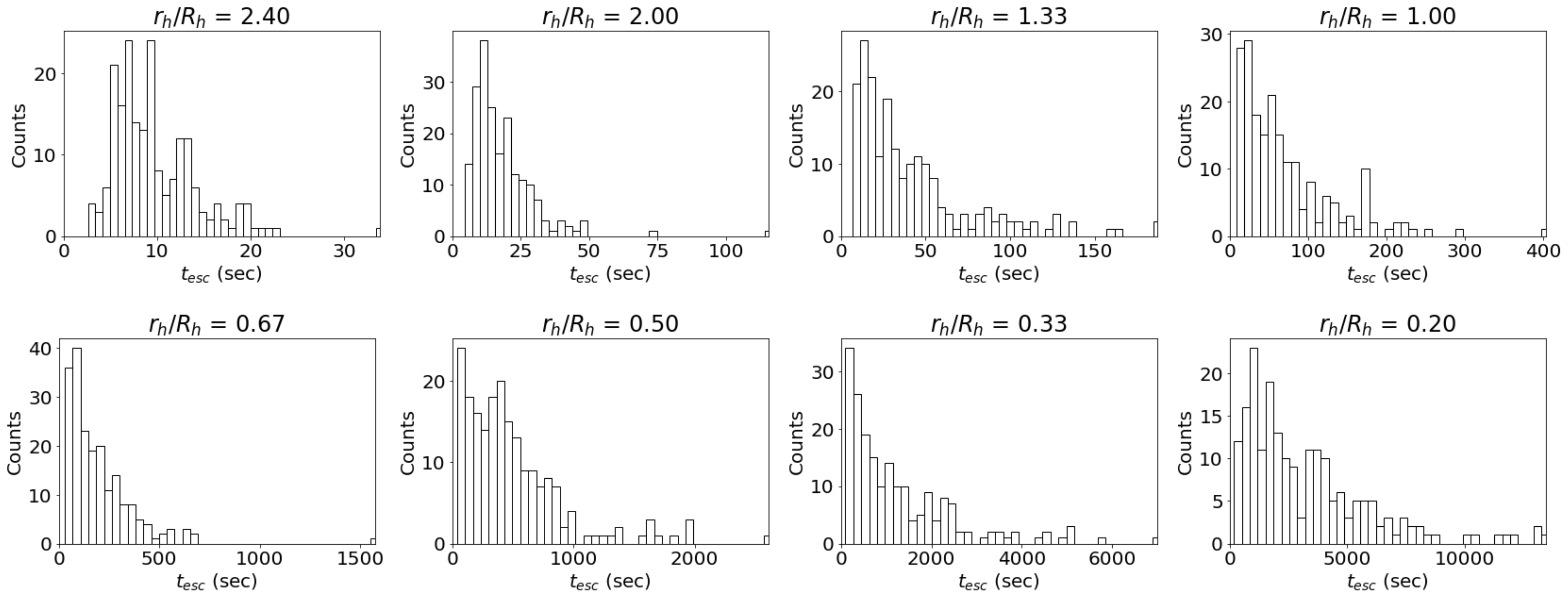}
\caption{\label{fig:SI_Esc}
\changed{\textbf{Escape time histograms for different relative sizes.}}
Histograms of the escape times $t_{esc}$ of a roller placed in the wake of an obstacle ([$x=R_h+r_h,y=0$]) for different \changed{relative sizes $r_h/R_h$} of the obstacle and Debye length $b/r_h=0.1$. }
\end{figure*}

\begin{figure*}
\centering
\includegraphics[width=\textwidth]{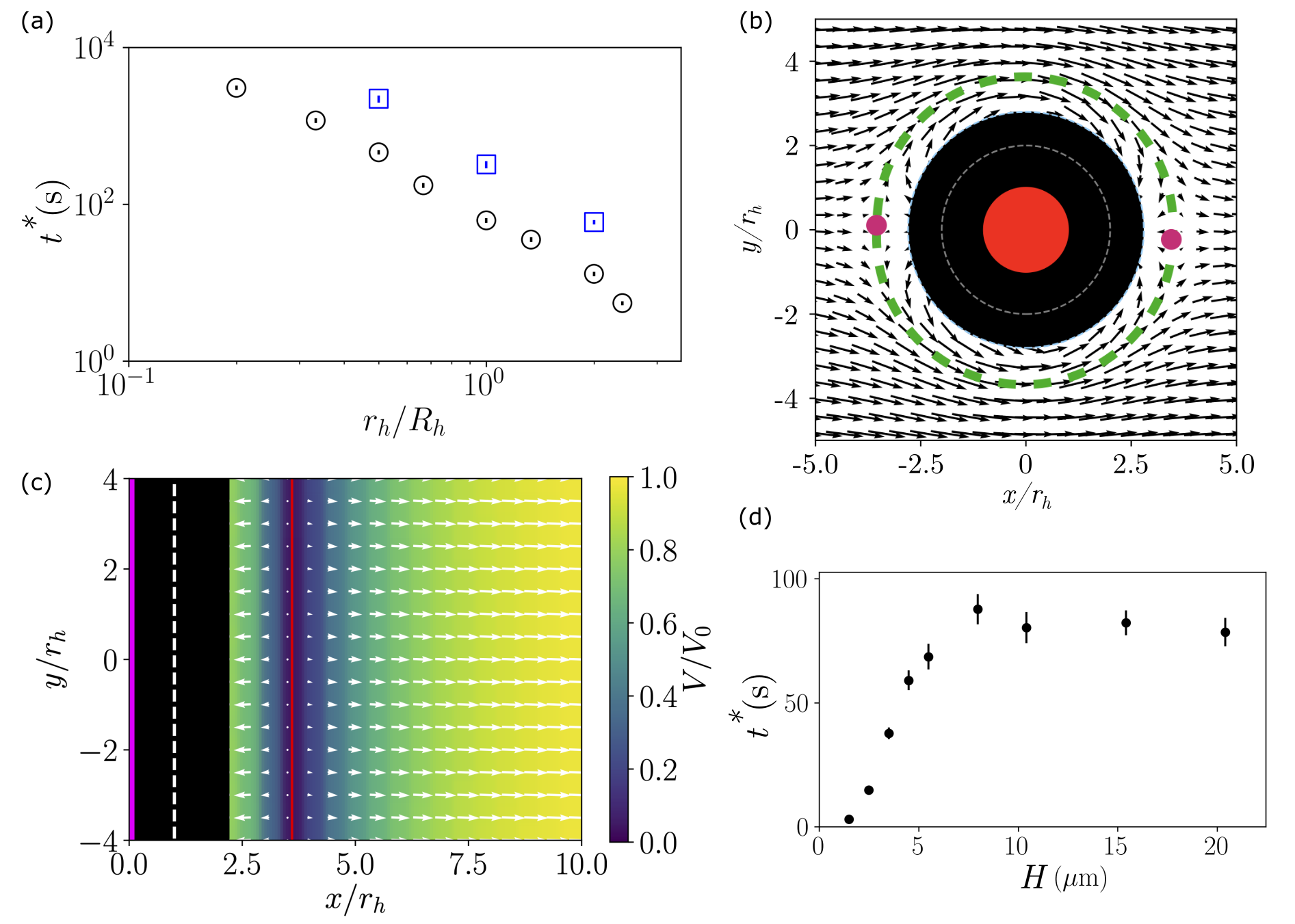}
\caption{\label{fig:SI_hght_deb_torq_res}
\changed{\textbf{Roller-obstacle interaction.}}
(a) Log-log plot of the mean escape time $t^*$ as a function of relative \changed{size} $r_h/R_h$ for different multiblob resolutions: 12 blobs (\protect\opencircle) and 42 blobs (\protect\opensquare) per microroller. 
Errors bars denote the standard error of the mean.
(b) Flow field of a microroller at $z/r_h = 1.29$ around a pillar ($r_h/R_h=1$) showing two \changed{saddle} points (purple) and a separatrix encircling the pillar (green dashed line). On the separatrix the flow converges for $x<0$ and diverges for $x>0$. The \changed{saddle} points are slightly shifted from $y=0$ due to the finite resolution of the pillar in the simulations.
(c) Microroller velocity field in the \textit{xy} plane calculated downstream of a wall ($r_h/R_h=0$, $H=5.5$ \textmu m, 10 \textmu m long). The microroller velocities are normalized to the microroller velocity in bulk $V_0$. The magenta line indicated the wall position, the white line denotes the position where the roller is in hard contact with the wall. The black area is drawn where the electrostatic repulsion dominates the dynamics of the roller. The red line indicates the \changed{saddle} line.
(d) The mean escape time $t^*$ as a function of pillar height $H$. Errors bars denote the standard error of the mean.}
\end{figure*}

\begin{figure*}
\centering
\includegraphics[width=0.8\textwidth]{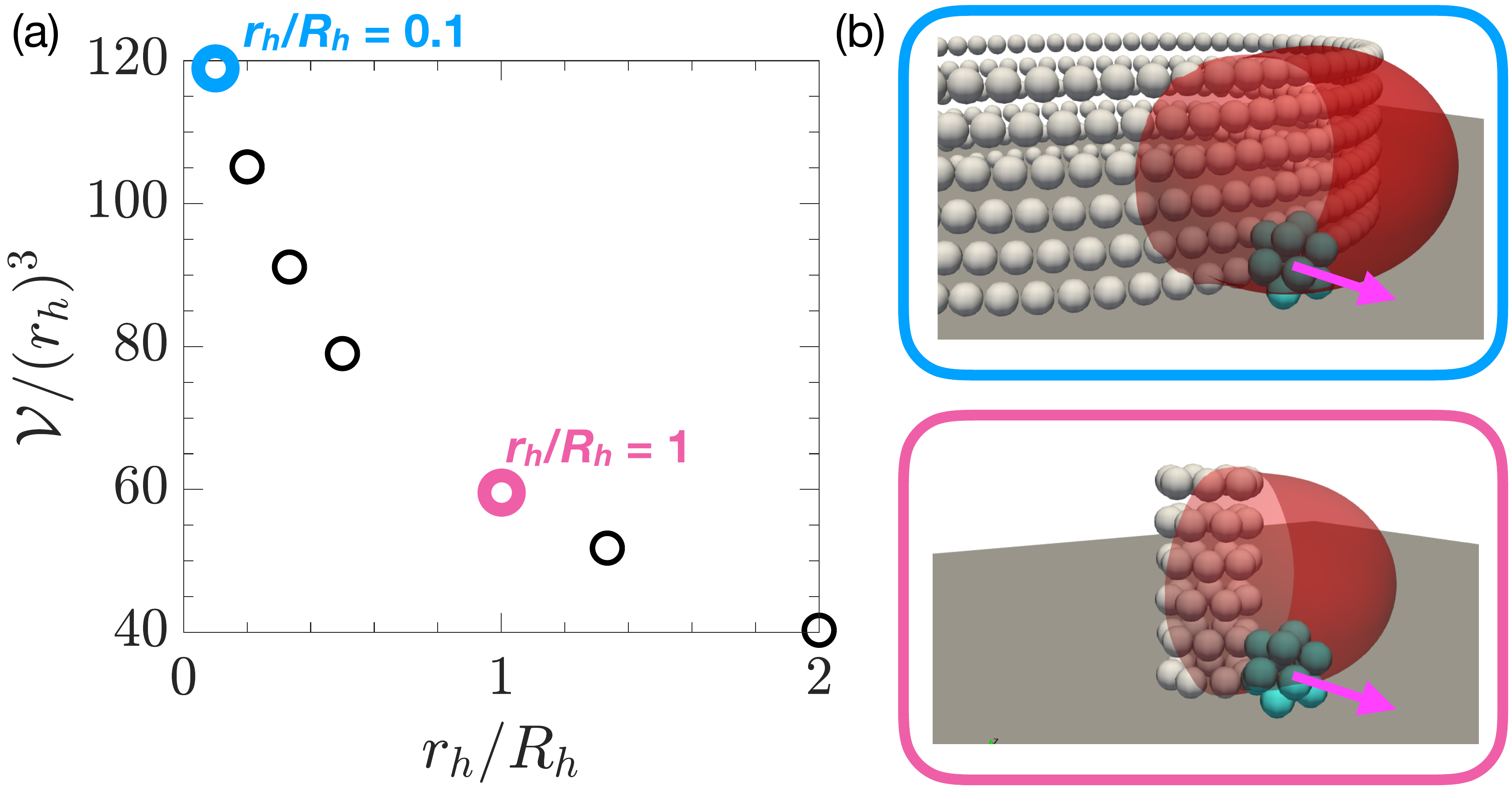}
\caption{
\changed{\textbf{Iso-surface volume in simulations as function of relative size.}}
(a) Volume $\mathcal{V}$ enclosed by the iso-surface $u_x = -V_0$ of the flow induced by the traction forces on the obstacle surface, when the microroller is at a separation distance $d_x=2r_h$, as a function of the relative \changed{size} $r_h/R_h$. The volume is restricted to the region ahead of the obstacle $x>x_{obs}+R_h$ (b) 3D representation of  $\mathcal{V}$ (red transparent contour) for $r_h/R_h = 0.1$ (top) and $r_h/R_h = 1$ (bottom).
\label{fig:SI_volume_iso_uOx}
}
\end{figure*}

\begin{figure*}
\centering
\includegraphics[width=0.65\textwidth]{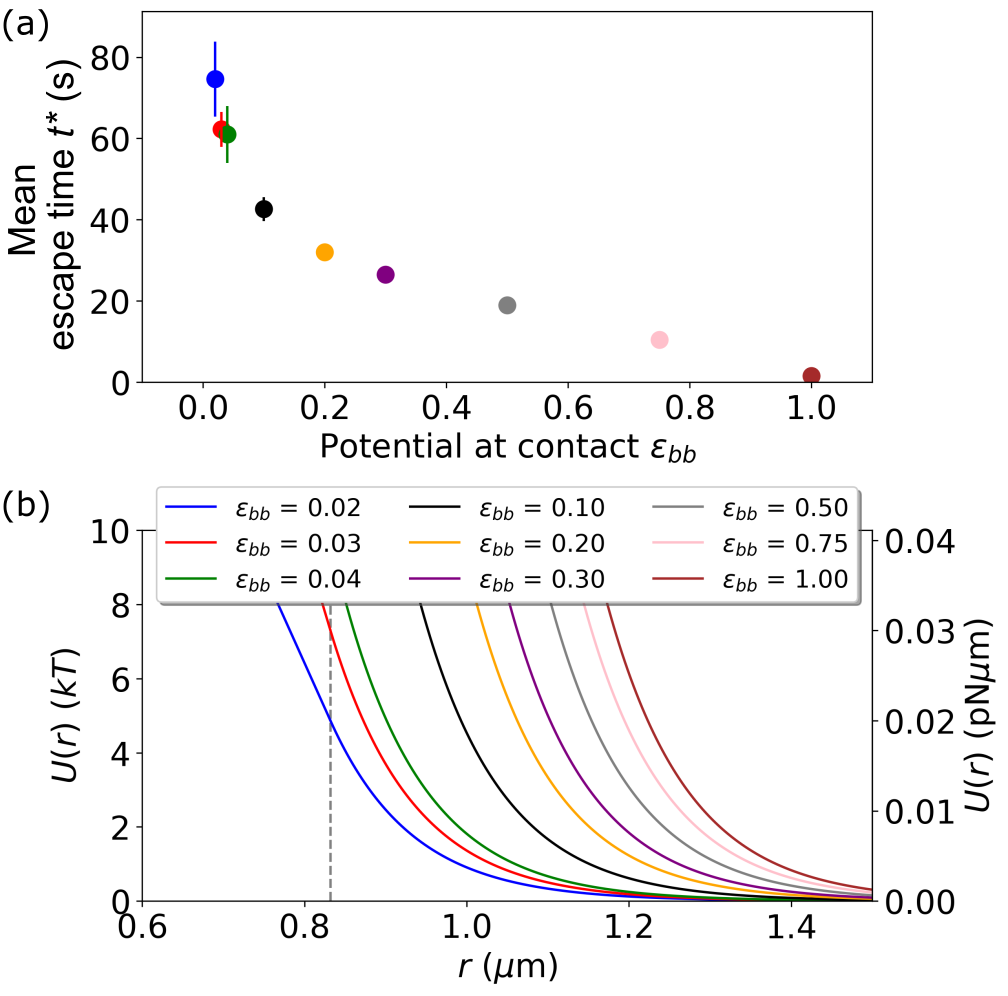}
\caption{
\label{fig:potentials}
\textbf{Influence of the microroller-to-obstacle interaction strength $\epsilon _{bb}$ on the trapping \changed{time}.}
(a) Mean escape time $t^*$ as a function of the potential of contact $\epsilon _{bb}$, with the error bar as the standard error of the mean, for $r_h/R_h=1$ and $(br_h)^{-1}=0.1$. 
(b) The blob-blob interaction potentials with different potentials at contact $\epsilon _{bb}$ with colors matching the data points in (a). The grey vertical dashed line denotes contact of the blobs.
}
\end{figure*}